\documentclass[aps,twocolumn]{revtex4}
\usepackage{graphicx}
\usepackage{dcolumn}
\newcommand{\karplus}[3]{\mbox{$^3J_{\text{#1\,#2}}(\text{#3})$}}
\newcommand{\jj}[2]{\mbox{$^3J_{\text{#1\,#2}}$}}
\newcommand{\jjpl}[2]{\mbox{$J_{\text{#1\,#2}}$}}
\newcommand{\jjnm}[1]{\mbox{$J_{\text{#1}}$}}
\newcommand{\A}{\mbox{$\Uparrow$}}
\newcommand{\J}{\mbox{$\cdot\,$}}
\begin{document}
\title{Measurability of side chain rotational isomer populations: \\
NMR and molecular mechanics of cobalt glycyl-leucine dipeptide model system}
\author{Christopher Haydock}
\email{haydock@appliednewscience.com}
\affiliation{Applied New Science LLC, Rochester, Minnesota 55901, USA}
\author{Nenad Jurani\'{c}}
\author{Franklyn G. Prendergast}
\author{Slobodan Macura}
\affiliation{Department of Biochemistry and Molecular Biology, \\
Mayo Clinic and Foundation, Rochester, Minnesota 55905, USA}
\author{Vladimir A. Liki\'{c}}
\affiliation{Bio21 Molecular Science and Biotechnology Institute, \\
University of Melbourne, VIC 3010, Australia}
%
%
\begin{abstract}
The cobalt glycyl-leucine dipeptide is a model system for studying the effects
of Karplus equation calibration, molecular mechanics accuracy, backbone
conformation, and thermal motions on the measurability of side chain rotational
isomer populations.  We analyze measurements of 8 vicinal coupling constants
about the $\alpha$ to $\beta$-carbon and $\beta$ to $\gamma$-carbon bonds of the
leucine side chain and of 10 NOESY cross relaxation rates across these bonds.
Molecular mechanics and peptide and protein crystallographic databases are an
essential part of this analysis because they independently suggest that the
trans gauche$^+$ and gauche$^-$ trans rotational isomers of the leucine side
chain predominate.  They also both suggest that puckering of the cobalt
dipeptide ring system reduces the gauche$^+$ gauche$^+$ rotational isomer
population to less than about 10\%.  At the present $\pm$1 Hz calibration
accuracy of Karplus equations for vicinal coupling constants, the predominant
trans gauche$^+$ and gauche$^-$ trans rotational isomer populations can be
measured with about 5\% accuracy, but the population of the gauche$^+$
gauche$^+$ rotational isomer is probably very near or just below the limit of
measurability.  These estimates also depend upon qualitative
assessments of the accuracy of the molecular mechanics energy wells.
We introduce gel graphics that are ideally suited to presenting
qualitative error and measurability estimates.
\end{abstract}
%
%
\maketitle
%

\section{INTRODUCTION}

Comparisons of NMR structures with X-ray structures show that
vicinal coupling constants accurately measure
the backbone torsion angles of proteins \cite{Cornilescu00,Sprangers00,Wang96}.
In the best X-ray structures multiple conformations of a particular
side chain can be confidently identified
\cite{Esposito00,Jelsch00,Yamano97,Burling96,Teeter93},
but population estimates may not be very accurate.
The populations of the three $\chi^1$ rotational isomers of protein
side chains can be determined from vicinal coupling constants if the
rotational isomers are assumed to have ideal staggered
conformations \cite{Hennig99,Karimi-Nejad94}.
If the side chain rotational isomers are not assumed to have ideal
conformations and the amplitudes of the torsion angle fluctuations
of each rotational isomer are also unknown,
then in general it is only possible to measure rotational isomer
populations as statistical averages over many side chains \cite{West98}.
Even with a very complete set of vicinal coupling constants
and NOESY cross relaxation rates it is difficult
to measure the populations of more than two rotational
isomers of a single side chain \cite{Dzakula92}.
Protein NMR and crystallographic data available now or in the foreseeable
future simply do not have the resolution to measure
the populations of all possible side chain rotational
isomers at anywhere near the 1\% level of accuracy.
Understanding protein properties such as fluorescence
intensity decay spectra
\cite{Sillen00,Moncrieffe00,Moncrieffe99,Antonini97,Verma96,Haydock93},
hydrogen ion association constants \cite{Onufriev00,Luo98,You95},
or global stability \cite{Vohnik98,Wilson95} often requires
that an NMR or X-ray structure be supplemented with molecular
mechanics structure calculations of the conformations
of a particular side chain.
An ideal structure determination method would incorporate
these supplemental molecular mechanics calculations and simultaneously
fit NMR or crystallographic data.
The consistency of the data and the incorporated molecular mechanics
could be judged by existing methods \cite{Kleywegt95,Brunger93}
for assessing when a model over-fits the data.
In the case of measuring side chain rotational isomer populations
it might be possible to measure the population of one
or two prominent rotational isomers and assess the
measurability of other molecular mechanically plausible rotational isomers.
In this work we take a simple first step in this direction
with an analysis of the cobalt glycyl-leucine dipeptide model system.
This model system has the advantages that
the vicinal coupling constants and NOESY cross relaxation rates
can be accurately measured on samples with natural isotope abundance
and that the cobalt dipeptide ring system restrains
the dipeptide backbone in a single conformation.

The two background sections give essential information about
the conformational analysis of leucine side chain rotamers
and about the accuracy of the Karplus equation coefficients.
The experimental section gives the
vicinal coupling constant and NOESY cross relaxation rate data
for the cobalt glycyl-leucine dipeptide.
Simple two and three rotational isomer models suggested by
conformational analysis are compared and fit to some of this data.
The computational results and discussion section examines
the molecular mechanics energy map, the effect of intramolecular
thermal motions on calculated NMR coupling constants and cross
relaxation rates, and the Monte Carlo probability density functions
of the rotational isomer populations.
For the simple cobalt dipeptide model system
these probability density functions confirm the preliminary
analysis in the experimental section and additionally
suggest that the populations of the remaining rotational
isomers are unmeasurable at present.
Our analysis shows that while the NMR data is not fit too well by the
simplest models neither does this data give any guidance in
selecting among the multitude of models with improved fits.

\section{BACKGROUND}

\subsection{Leucine side chain conformational analysis}

The leucine side chains of crystallographic oligopeptide and protein
structures strongly prefer the trans guache$^+$ and gauche$^-$ trans
rotational isomers \cite{Benedetti83,Schrauber93}.
Conformational analysis predicts the backbone-dependent
stability of protein side chain conformations
and explains the rotamer preferences observed in high resolution
crystallographic structures of proteins \cite{Dunbrack94,Dunbrack93}.
This analysis is equally applicable to the leucine side chain
of the cobalt glycyl-leucine dipeptide and leads to the same conclusions
found for leucine side chains in proteins.
In either case the predominant rotational isomers of the
leucine side chain are trans guache$^+$ and gauche$^-$ trans.
The butane and {\em syn}-pentane effects are known from the
conformational analysis of the simple hydrocarbons $n$-butane and $n$-pentane.
The {\em syn}-pentane conformations, which are guache$^+$ guache$^-$
or guache$^-$ guache$^+$, are about 3.3 kcal/mol higher in energy than
the extended trans trans conformation.
A molecular conformation is said to be destabilized
by the {\em syn}-pentane effect when one (or more) five atom fragments
of the molecule are in {\em syn}-pentane like conformations.
The conformational analysis of a peptide or protein identifies
unfavorable side chain conformations primarily by searching for
{\em syn}-pentane effects among all possible $n$-pentane
fragments with the C$^{\alpha}$--C$^{\beta}$ bond
in the second or third fragment position.
When the C$^{\alpha}$--C$^{\beta}$ and C$^{\beta}$--C$^{\gamma}$ bonds
are in the second and third fragment positions,
the analysis gives backbone-independent rotamer preferences
because the fragment conformation does not depend on the backbone
$\phi$ and $\psi$ angles.
When the backbone N--C$^{\alpha}$ or C$^{\prime}$--C$^{\alpha}$ bond is
in the second and the C$^{\alpha}$--C$^{\beta}$ bond
is in the third fragment position,
the analysis gives backbone-dependent rotamer preferences.
For a leucine residue there are eight pentane fragments to consider:
four backbone-independent fragments of the pattern
\newcommand{\alphgamma}{C$^{\alpha}$--C$^{\beta}$--C$^{\gamma}$}
\newcommand{\alphgamith}{C$_i^{\alpha}$--C$_i^{\beta}$--C$_i^{\gamma}$}
(N,C$^{\prime}$)--\alphgamma--(C$^{\delta 1}$,C$^{\delta 2}$),
and four backbone-dependent fragments of the patterns
(C$_{i-1}$,O-$\!$ -$\!$ -HN$_i$)--N$_i$--\alphgamith
or (N$_{i+1}$,O$_i$)--C$_i^{\prime}$--\alphgamith,
where leucine is the $i$th residue and O-$\!$ -$\!$ -HN$_i$
is an assumed hydrogen bond acceptor of HN$_i$.
A very clear Newman projection diagram of these eight
fragments is shown in Fig.\ 2 of Ref.\ \onlinecite{Dunbrack94}.
Of the nine possible leucine rotamers only two,
trans guache$^+$ and gauche$^-$ trans, have no backbone-independent
{\em syn}-pentane interactions, six have one interaction,
and one, guache$^+$ gauche$^-$, has two such interactions.
Because the identical backbone-independent fragments are present in the
cobalt dipeptide and proteins,
the backbone-independent conformational analysis for proteins
applies equally to the cobalt dipeptide.

\begin{figure}
\resizebox{8.3cm}{!}{\includegraphics{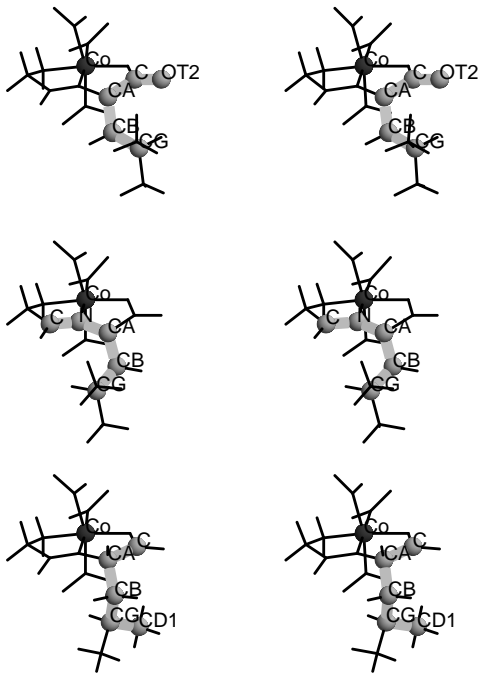}}
\caption{Stereo views of three of the nine possible
rotational isomers of the cobalt dipeptide leucine side chain:
{\em top}, trans gauche$^+$; {\em middle}, gauche$^-$ trans; and
{\em bottom}, gauche$^+$ gauche$^+$.
The three rotational isomers each have a single {\em syn}-pentane
interaction, which occurs in the pentane fragment shown as a grey ball
and stick representation with exagerated stick diameters and labeled atoms.
The cobalt atom is the labeled central black sphere.
The gauche$^+$ trans rotational isomer (not shown) also has only
one {\em syn}-pentane interaction, but turns out to have
a slightly higher energy than the gauche$^+$ gauche$^+$ rotational isomer.
The five remaining rotational isomers (not shown) all have two 
{\em syn}-pentane interactions and are several kcal/mol higher in energy.
Note that the {\em syn}-pentane interactions of the trans gauche$^+$ and
gauche$^-$ trans rotational isomers ({\em top} and {\em middle})
depend on $\chi^1$ and in addition the leucine backbone $\psi$ or
$\phi$ torsion angles, respectively.  These are thus
backbone-dependent {\em syn}-pentane interactions.
The {\em syn}-pentane interaction of the gauche$^+$ gauche$^+$
rotational isomer ({\em bottom}) depends on the $\chi^1$
and $\chi^2$ torsion angles and is a backbone-independent.
The trans gauche$^+$ and gauche$^-$ trans rotational isomers
have the lowest energies because the backbone-dependent
{\em syn}-pentane interactions can be relieved by chelate
ring puckering.}
\label{fig1:syn_pentane}
\end{figure}

The conformational analysis of backbone-dependent rotamer preferences
is important because the cobalt dipeptide backbone
forms two approximately planar chelate rings with
$\phi$ and $\psi$ angles differ somewhat from
the angles most commonly found in protein structures.
To apply backbone-dependent conformational analysis to the leucine
side chain of the cobalt dipeptide (Fig.\ \ref{fig1:syn_pentane})
simply note that leucine is the
second residue and substitute the atom names Co, O$^{t1}$, and O$^{t2}$
for the names O-$\!$ -$\!$ -HN$_i$, O$_i$, and N$_{i+1}$ in the
above five atom fragments, where O$^{t1}$ is the terminal carboxyl
oxygen bonded to cobalt and O$^{t2}$ is the uncomplexed carboxyl oxygen.
(The reverse substitution of O$^{t1}$ and O$^{t2}$ inconveniently
generates the dipeptide from a protein with a very unlikely backbone
conformation that has colliding amide groups.)
With this identification the leucine backbone torsion angles
$\phi_2$ and $\psi_2$ are both near 180 degrees.
If $\phi_2 = -174.74$ degrees and leucine $\chi^1$ is gauche$^-$,
there is a {\em syn}-pentane interaction between
the glycine carbonyl carbon and the leucine C$^{\gamma}$
(Fig.\ \ref{fig1:syn_pentane} {\em middle}).
If $\psi_2 = 174.74$ degrees and leucine $\chi^1$ is trans,
there is a {\em syn}-pentane interaction between
the uncomplexed terminal oxygen and the leucine C$^{\gamma}$
(Fig.\ \ref{fig1:syn_pentane} {\em top}).
Because $\phi_2$ and $\psi_2$ of the cobalt dipeptide are both near
these critical angles, {\em syn}-pentane effects could destabilize
both of the leucine side chain trans guache$^+$ and gauche$^-$ trans
rotamers, which are observed in most crystallographic structures
and preferred by the backbone-independent conformational analysis.

Crystallographic studies of copper and cobalt dipeptides
and the vicinal coupling constants about both N--C$^{\alpha}$
bonds of the cobalt glycyl-leucine dipeptide suggest that
the $\phi_2$ and $\psi_2$ torsion angles of the cobalt glycyl-leucine
dipeptide depart from 180 degrees by as much as 10 or 20 degrees.
In the crystallographic structures of cobalt glycyl-glycine dipeptides
\cite{Prelesnik84} and copper dipeptides \cite{Freeman77}
the chelate ring conformations vary over a wide range.
The peptide backbone atoms are typically displaced by up to 0.1 or 0.2
angstrom from the mean plane of the chelate rings and 
the angles between 3-atom segments of the chelate rings are typically
5 or 10 or perhaps as large as 20 degrees.
Both these measures of chelate ring puckering imply backbone torsion angles
of 180$\pm$10 degrees with maximum deviations from 180 degrees of no more
than about 20 degrees.  The variability of the chelate ring conformations
is thought to arise from intermolecular contacts within the crystall.
Even if a crystallographic structure of the cobalt dipeptide were available,
no reliable predictions about the solution conformation of the chelate rings
could be made because the conformational distortions caused by intermolecular
contacts are not well understood.
In various DMSO plus D$_2$O mixtures measurements of the four
H--N--C$^{\alpha}$--H vicinal coupling constants about the glycine
N--C$^{\alpha}$ bond of the cobalt glycyl-leucine dipeptide show a rotation
about this bond of $-$10 to $-$20 degrees \cite{Juranic93}.
This rotation angle is relative to an eclipsed substituent atom geometry
about the N--C$^{\alpha}$ bond and implies a puckered amino-peptidato
chelate ring.  Though this puckering gives no direct information about
the leuine $\phi_2$ and $\psi_2$ torsion angles, the presence of puckering
in solution shows that the intermolecular contacts within a crystall are not
the only cause of puckering.
The C$^{\prime}$--N--C$^{\alpha}$--H vicinal coupling constant
about the N--C$^{\alpha}$ bond of the cobalt glycyl-leucine dipeptide
is about 0.3 Hz larger than the same coupling constant of the
cobalt glycyl-glycine dipeptide \cite{Juranic93}.
(In Ref.\ \onlinecite{Juranic93} the first atom of this coupling
constant is mistakenly labeled C$^{\alpha}$ rather than C$^{\prime}$.)
This coupling constant difference implies that the cobalt dipeptide
$\phi_2$ torsion angle is about $-170\pm10$ degrees.

Simple inspection of backbone-dependent rotamer libraries
suggests that a 20 degree departure from the
backbone torsion angles of planar chelate rings
can diminish or even eliminate the gauche$^+$ $\chi^1$ rotamer
preference of the leucine side chain.
The proportions of gauche$^-$, gauche$^+$, and trans $\chi^1$
side chain rotamers change fairly dramatically
for both 30 and 20 degree backbone
angle increments within the backbone-dependent rotamer libraries
of Dunbrack and Karplus \cite{Dunbrack94,Dunbrack93}.
The rotamer libraries seem to show that backbone-independent interactions
are slightly more important than backbone-dependent interactions.
In the backbone-dependent rotamer library for straight
side chains \cite{Dunbrack94}, which does not include leucine,
the gauche$^-$ and trans $\chi^1$ rotamers appear with
a combined frequency of 20 to 30\% in the
two 30 by 30 degree square cells adjacent to the point $\phi,\psi = 180$
degrees, even though backbone-dependent {\em syn}-pentane interactions
favor gauche$^+$ $\chi^1$ rotamers.
A similar trend seems to hold throughout the
backbone-dependent rotamer library, $\chi^1$ rotamers
excluded by {\em syn}-pentane interactions appear with diminished
probability rather than being completely excluded.
In contrast, the backbone-independent {\em syn}-pentane interactions
of the leucine side chain exclude the gauche$^+$ $\chi^1$ rotamer
with striking completeness.
In crystallographic structures on which the rotamer library is based less
than 2\% of the leucines are gauche$^+$ $\chi^1$ rotamers \cite{Dunbrack94}.
The exclusion of leucine gauche$^+$ $\chi^1$ rotamers as well
as gauche$^-$ $\chi^2$ rotamers seems to hold over all known
classes of protein backbone structures \cite{Schrauber93}.
The crystallographic and NMR evidence cited in the last paragraph
suggests that the $\phi_2$ and $\psi_2$ torsion angles of the cobalt
dipeptide both depart from 180 degrees by as much as 10 or 20 degrees.
This is more than enough to relieve the backbone-dependent
{\em syn}-pentane interactions with the leucine C$^{\gamma}$
and diminish the backbone-dependent preference
for gauche$^+$ $\chi^1$ rotamers of the leucine side chain.

\subsection{Accuracy of Karplus equation calibration}

Theoretical calculations show that the H--C--C$^{\prime}$--H$^{\prime}$
vicinal proton coupling constant depends on the torsion angle
$\phi$(H,C,C$^{\prime}$,H$^{\prime}$) around the C--C$^{\prime}$ bond,
the electronegativity and orientation of substituent groups
on the C and C$^{\prime}$ atoms,
the bond angles $\theta$(H,C,C$^{\prime}$)
and $\theta$(C,C$^{\prime}$,H$^{\prime}$),
and the length of the C--C$^{\prime}$ bond \cite{Karplus63}.
The same symbol $\phi$ serves here for the torsion angle
between the vicinally coupled spins and elsewhere for the protein backbone
torsion angle, but the meaning should always be clear from the context.
The theoretical dependence of the coupling constant on
the torsion angle is approximated by a Karplus equation,
which is often written in the form
\begin{equation}
  ^3J(\phi) = A\cos^2\phi - B\cos\phi + C,
\end{equation}
where $\phi$ is the torsion angle around the C--C$^{\prime}$ bond.
For peptide side chain vicinal coupling constants the torsion angle $\phi$
is equal to the side chain torsion angle to within a phase shift
that is approximately an integer multiple of 120 degrees.
The Karplus equation for a H--C--C$^{\prime}$--C$^{\prime\prime}$
heteronuclear vicinal coupling constant also has the same form,
where the torsion angle is now $\phi$(H,C,C$^{\prime}$,C$^{\prime\prime}$)
around the same C--C$^{\prime}$ bond.
The accuracy of theoretical coupling constant calculations or of the
Karplus equation fit to such calculations is at best around $\pm$1 Hz.
A more accurate Karplus equation is obtained by adjusting
the coefficients to fit experimental coupling constant data.
The greatest improvement occurs when the Karplus equation
coefficients are calibrated for a four atom fragment with
specific functional groups substituted in a specific orientation
on the central two atoms.
For example, the error in the coupling constants predicted
by the Karplus equation calibrated for the protein
H--N--C$^{\alpha}$--C$^{\beta}$ heteronuclear coupling constant
is $\pm$0.25 Hz, as estimated by the RMS difference between
the Karplus curve and the fit experimental data \cite{Wang96}.
If a Karplus equation that is calibrated for a four atom fragment
with specific substituent groups is applied to the same four atom fragment
with different substituent group chemistry or orientation,
then the errors in the predicted coupling constants are dramatically
increased.  Similar large errors in the predicted coupling constants
result if the Karplus equation is fit
to a collection of experimental coupling constants
that are all measured for the same fixed four atom fragment,
but with a variety of functional groups
substituted in a variety of orientations on the two central atoms.
For one data set of over 300 experimental measurements
of the H--C--C$^{\prime}$--H$^{\prime}$ coupling constant
in about 100 conformationally rigid compounds, which are largely
6-membered rings with holding groups, the RMS difference between
the fit Karplus curve and experimental data is 1.2 Hz \cite{Haasnoot80}.
In this data set carbon and oxygen are the most frequent
substituent atoms bonded to the central two carbon atoms of the
H--C--C$^{\prime}$--H$^{\prime}$ fragment,
while nitrogen, sulfur, halogen, silicon, and selenium substituent atoms
occur in smaller numbers.
These two examples span the accuracy range
of most calibrated Karplus equations, that is, errors in predicted
coupling constants are in the range $\pm$0.25 to $\pm$1 Hz.
For a specific substituent group chemistry and orientation
the error may be around $\pm$0.5 Hz or even as low as
$\pm$0.25 Hz in favorable cases.
For small to moderate variations in
substituent group chemistry or orientation the error
probably is in the range $\pm$0.5 Hz to $\pm$1 Hz.

Studies of the vicinal coupling constants of peptides
and closely related compounds suggest that the above generalizations
about the accuracy of calibrated Karplus equations apply to the
vicinal coupling constants about the C$^{\alpha}$--C$^{\beta}$ and
C$^{\beta}$--C$^{\gamma}$ bonds of the cobalt dipeptide leucine side chain.
Simple information about the effect
of substituent group chemistry comes from alanine and its analogues,
which have a single H--C$^{\alpha}$--C$^{\beta}$--H coupling constant because
of the three-fold symmetry of the methyl side chain.
The experimental H--C--C--H coupling constant
of ethane is 8.0 Hz \cite{BothnerBy65}, the H--C$^{\alpha}$--C$^{\beta}$--H
coupling constant of the alanine dipeptide is 7.3 Hz \cite{Kopple73},
and the H--C$^{\alpha}$--C$^{\beta}$--H coupling constant
of the amino acid alanine remains almost exactly 7.3 Hz
over the pH range of 0.5 to 12.5 \cite{Roberts70}.
Replacing two protons on one ethane carbon atom with one carbon
and one nitrogen substituent group drops the coupling constant by 0.7 Hz;
changing the nitrogen substituent group electronegativities
through the range ammonium $>$ acetamide $>$ amide and changing the carbon
substituent group through the range carboxyl $>$ N-methylamide $>$ carboxylate
does not change the coupling constant at all.
But there are many counter examples to this seeming insensitivity
to substituent group chemistry.
The H--C--C--H coupling constant of propane is 7.3 Hz
and of isopropylamine is 6.3 Hz \cite{BothnerBy65}.
Replacing one ethane proton with one carbon substituent group already drops
the coupling constant to the value observed for alanine,
which has an additional nitrogen substituent group,
and replacing a second proton on the same carbon atom
with a nitrogen substituent group, making the substituted carbon
equivalent to the alanine $\alpha$-carbon,
drops the coupling constant by an additional 1.0 Hz.
The H--C$^{\alpha}$--C$^{\beta}$--H coupling constants
of various alanine dipeptide derivatives are in the range
6.9 to 7.3 Hz \cite{Kopple73},
which shows that even substituent group changes one peptide bond
removed from the $\alpha$-carbon can change the coupling
constant about the C$^{\alpha}$--C$^{\beta}$ bond by at least 0.4 Hz.

The $\beta$-carbon atoms of all the other amino acids lack
the three-fold symmetry of the alanine $\beta$-carbon.
The above examples of alanine mehtyl proton coupling
across the C$^{\alpha}$--C$^{\beta}$ bond
probably underestimate the coupling constant variation
due to $\alpha$-carbon substituent group chemistry
and totally ignore the effect of $\beta$-carbon substitution.
For leucine two H--C$^{\alpha}$--C$^{\beta}$--H coupling constants
between the $\alpha$ and $\beta$-protons and four heteronuclear
coupling constants between the amide nitrogen and carbonyl carbon
and $\beta$-protons are usually measurable.
The simplest models for these vicinal coupling constants
about the C$^{\alpha}$--C$^{\beta}$ bond assume
ideal gauche or trans torsion angles between the coupled spins
and have four parameters: the populations of two of the three
$\chi^1$ rotational isomers and the gauche and trans coupling constants.
The heteronuclear N--C$^{\alpha}$--C$^{\beta}$--H trans coupling
constant of the leucine cation apparently decreases
by 0.6 Hz when the cation is converted into the anion \cite{Fischman78}.
The effect of $\alpha$-carbon substituent chemistry
on the coupling constants about the C$^{\alpha}$--C$^{\beta}$ bond
is also seen in the 1-substituted derivatives of 3,3-dimethylbutane,
which are analogues of the amino acid leucine
with the $\alpha$-carbon and side chain intact
and with various replacements for the amine and carboxylate groups.
Both the gauche and trans coupling constants of these
analogues vary over the range of 0.7 Hz \cite{Whitesides67}.
Furthermore, this same study found a 1 Hz difference
in the average gauche coupling constant depending on whether
the 1-substituent was gauche or trans to the coupled proton
on the second carbon.  This suggests that two separate Karplus
equations are required for the two $\beta$-protons of leucine.
Substituent orientation effects require two different
Karplus equations for predicting
the $\beta$-proton coupling constants of proline \cite{Madi90}.
In a similar way the electronegativity corrections
to the \karplus{H$\alpha$}{H$\beta$}{$\phi$} Karplus equation for leucine
probably depend on the orientation
of the substituent groups with respect to both the H$^{\alpha}$
and H$^{\beta}$ protons \cite{Bystrov76}.

The existing calibrations of the Karplus equations for vicinal
coupling about the C$^{\alpha}$--C$^{\beta}$ bond suffer from
several sources of error.
Most assume that for each $\alpha$-carbon bonded atom one Karplus
equation predicts the coupling of this atom to both $\beta$-protons.
The calibrations are done with sets of model compounds that
have normal or sometimes rather far from normal peptide backbone chemistries
and that have a range of standard and nonstandard amino acid side chains.
Errors arise because the set of model compounds is too varied or
because none of the model compounds closely match the molecule of interest,
whether it be a protein or as in this study the cobalt dipeptide.
Model compounds such as
2,3-substituted bicyclo[2.2.2]octanes \cite{Fischman80,Kopple73},
gallichrome \cite{DeMarco78},
$\alpha$-amide-$\gamma$-butyrolactones \cite{Cung82},
differ from standard proteins in both backbone and side chain structure.
The match to the molecule of interest may depend on a choice of
coupling constants about several different bonds of the model compound.
Because the gallichrome backbone and ornithyl side chains out to the
$\beta$-carbon have essentially the same structure as standard proteins,
the C$^{\alpha}$--C$^{\beta}$ bonds of gallichrome are fairly
well matched to the C$^{\alpha}$--C$^{\beta}$ bonds of proteins.
The substituent groups on the $\beta$ and $\gamma$-carbons
are obviously quite different from those on the $\alpha$-carbon
and coupling about the C$^{\beta}$--C$^{\gamma}$ and
C$^{\gamma}$--C$^{\delta}$ bonds is somewhat different from that
about the C$^{\alpha}$--C$^{\beta}$ bond.
If coupling constants about all three bonds are chosen
to calibrate the Karplus equation for coupling about the
C$^{\alpha}$--C$^{\beta}$ bond,
then the gallichrome model compound is not a very good match to proteins.
Model compounds such as
cyclo(triprolyl) peptide \cite{Kopple73},
an asparaginamide dipeptide,
oxytocin cystine-1 and 6,
alumicrocin \cite{Cung82},
match the backbone of standard proteins,
but the side chains may differ from a specific amino acid side chain
of interest.

The residuals between the Karplus curve and the calibration data
set are perhaps the best available indication of the accuracy
of a Karplus equation calibration.
However, these residuals generally underestimate the errors that occur
when the Karplus equation is then applied to predict
the vicinal coupling constants of a particular side chain of interest.
Fischman et al.\ calibrate the \karplus{C$^{\prime}$}{H$\beta$}{$\phi$}
Karplus equation by fitting the three Karplus coefficients to 4
coupling constants measured on two bicyclo-octanes \cite{Fischman78}.
For this fit the RMS residual per degree of freedom is 0.13 Hz.
Due to the small calibration data set this tiny observed residual
is a completely unreliable estimate of the true residual.
Kopple et al. calibrate the \karplus{H$\alpha$}{H$\beta$}{$\phi$}
Karplus equation by fitting to 10 coupling constants measured on
seven model compounds \cite{Kopple73}.
For this fit the RMS residual per degree of freedom is 0.47 Hz.
The data set is just large enough to give a reliable
residual estimate, but as discussed in the previous paragraph
the model compounds may not be a very good match to proteins.
DeMarco et al. calibrate the \karplus{H$\alpha$}{H$\beta$}{$\phi$}
Karplus equation by fitting 30 coupling constants measured about the
C$^{\alpha}$--C$^{\beta}$, C$^{\beta}$--C$^{\gamma}$ and
C$^{\gamma}$--C$^{\delta}$ bonds of the ornithyl
side chains of gallichrome \cite{DeMarco78}.
For this fit the RMS residual per degree of freedom is 0.92 Hz.
This fairly large residual is apparently the result of fitting
the coupling constants about all three side chain bonds.
The errors in this calibration may be even larger than $\pm$1 Hz
because fitting the coupling constants about all three side chain bonds
makes the gallichrome model compound a poor match to the
C$^{\alpha}$--C$^{\beta}$ bond of proteins
and may add additional bias error to that suggested by the residuals.
Cung and Marraud calibrate the \karplus{H$\alpha$}{H$\beta$}{$\phi$}
Karplus equation by fitting the three Karplus coefficients
and eight angle parameters to 16 coupling constants measured on
five model compounds \cite{Cung82}.
For this fit the RMS residual per degree of freedom is 0.49 Hz.
Note Cung and Marraud arrive at a standard deviation of half this value
by computing a straight RMS average of the 16 residuals
rather than by averaging over the 5 degrees of freedom actually present.
Though the five model compounds used for this calibration
are well matched to the C$^{\alpha}$--C$^{\beta}$ bond of proteins,
the errors of this calibration are likely to be substantially larger
than suggested by the residuals because the model compound
torsion angles are not estimated from crystallographic structures
or by molecular mechanics calculations.

Considering the wide variety of model compounds, the single calibration
for both $\beta$-protons, the observed residuals, and uncertainties
in the model compound structures,
it seems extremely unlikely that the errors in the
calibration of the Karplus equations for vicinal coupling about
the C$^{\alpha}$--C$^{\beta}$ and C$^{\beta}$--C$^{\gamma}$ bonds
are significantly less than $\pm$1 Hz, whether the molecule of
interest is a standard protein or peptide or as here the
glycyl-leucine dipeptide complexed to cobalt.

\section{EXPERIMENTAL RESULTS}

\begin{table}
\caption{\label{tab1:assignments}Proton assignments.  Shifts are in ppm.
The atom H$^{\beta 1}$ and atoms H$^{\delta 1}$ are {\em pro-R}
and the atom H$^{\beta 2}$ and atoms H$^{\delta 2}$ are {\em pro-S}.}
\begin{ruledtabular}
\begin{tabular}{l@{\extracolsep{36pt}}clc}
&$\delta$& &$\delta$\\ \hline
H$^{\alpha}$   & 4.1710 & H$^{\gamma}$   & 1.6582 \\
H$^{\beta 1}$  & 1.8125 & H$^{\delta 1}$ & 0.8700 \\
H$^{\beta 2}$  & 1.5440 & H$^{\delta 2}$ & 0.8200 \\
\end{tabular}
\end{ruledtabular}
\end{table}

The proton assignments in Table \ref{tab1:assignments} are model dependent.
These assignments depend on our assumption that the population
of the leucine side chain rotational isomers with
a gauche$^+$ $\chi^1$ torsion angle is small compared
to the population of rotational isomers with
gauche$^-$ and trans $\chi^1$ torsion angles.
Without this assumption an unambiguous assignment is not possible.
The conventional approach to assigning the $\beta$-protons
examines the vicinal coupling constants about the
C$^{\alpha}$--C$^{\beta}$ bond and exploits the fact that
a weak coupling is expected
for synclinal spins and a strong coupling for antiperiplanar spins.
When the leucine side chain $\chi^1$ torsion is gauche$^-$
the atoms H$^{\alpha}$ and H$^{\beta 1}$ are antiperiplanar
and the atoms H$^{\alpha}$ and H$^{\beta 2}$ are synclinal and
when $\chi^1$ is trans these angular magnitudes are reversed,
that is H$^{\beta 1}$ is synclinal and H$^{\beta 2}$
is antiperiplanar \cite{Kessler87}.
Thus the \jj{H$\alpha$}{H$\beta$} coupling constant does not
help with H$^{\beta}$ assignment, but is very helpful in determining
the ratio of gauche$^-$ to trans populations once this assignment is known.
The alternating synclinal antiperiplanar geometries of the
gauche$^-$ and trans rotational isomers produces a conjugate
\jj{H$\alpha$}{H$\beta$} coupling pattern that is diagnostic
of the absence of gauche$^+$ $\chi^1$ rotational isomers.
The average of the two \jj{H$\alpha$}{H$\beta$} couplings
is independent of the rotational isomer populations and the coupling ratio
(\jjpl{H$\alpha$}{H$\beta$1}$-$\jjnm{sc})/%
(\jjpl{H$\alpha$}{H$\beta$2}$-$\jjnm{sc})
is equal to the gauche$^-$ over trans population ratio.
For ideal geometries the Karplus equation predicts that
the synclinal coupling \jjnm{sc} = $A/4-B/2+C$
and that the average coupling is $5A/8+B/4+C$.
A very similar situation occurs with the \jj{H$\beta$}{H$\gamma$}
coupling constant.
Leucine side chain rotational isomers with a gauche$^-$ $\chi^2$ 
torsion angle are virtually excluded by backbone-independent
{\em syn}-pentane interactions \cite{Dunbrack94}.
When the $\chi^2$ torsion angle is gauche$^+$
the atoms H$^{\beta 1}$ and H$^{\gamma}$ are antiperiplanar
and the atoms H$^{\beta 2}$ and H$^{\gamma}$ are synclinal
and when $\chi^2$ is trans these angular magnitudes are reversed.
This coupling is again helpful with populations but not assignments
and the same conjugate coupling pattern is now diagnostic of the
absence of gauche$^-$ $\chi^2$ rotational isomers.
As noted in the introduction the $\chi^1$ and $\chi^2$ torsion
angles are highly correlated.  With only the trans gauche$^+$
and gauche$^-$ trans isomers populated trans $\chi^1$ implies
gauche$^+$ $\chi^2$ and gauche$^-$ $\chi^1$ implies trans $\chi^2$.
This correlation produces a doubly conjugate
\jj{H$\alpha$}{H$\beta$} and \jj{H$\beta$}{H$\gamma$}
correlation pattern.
When the trans gauche$^+$ rotational isomer predominates
the $\beta$1-proton couples weakly to the $\alpha$-proton and strongly
to the $\gamma$-proton and the $\beta$2-proton couples strongly
to the $\alpha$-proton and weakly to the $\gamma$-proton.
When the gauche$^-$ trans isomer predominates these coupling strengths
are all reversed.
If either one of these leucine side chain rotational isomers
predominates then the $\beta$-proton assignment can be made by inspection
of the \jj{C$^{\prime}$}{H$\beta$} heteronuclear coupling constants
\cite{Kessler87} because the leucine carboxyl carbon
and both $\beta$-protons are synclinal when the leucine side chain
$\chi^1$ torsion is gauche$^-$ and only the carboxyl carbon
and $\beta$2-proton are synclinal when $\chi^1$ is trans.

\begin{table}
\caption{\label{tab2:NMRdata}Experimental NMR data.
The NOESY cross relaxation rate units are s$^{-1}$
and the vicinal coupling constant units are Hz.
The standard deviations are estimated as described
in the experimental methods section.
A plus rather than plus and minus sign is placed between
a zero value and standard deviation solely to indicate
the fact that the measured quantities are theoretically nonnegative.}
\begin{ruledtabular}
\begin{tabular}{l@{\extracolsep{72pt}}l@{\extracolsep{6pt}}ll}
R$_{\text{H$\alpha$\,H$\beta$1}}$        & 0.038 &$\pm$& 0.010 \\
R$_{\text{H$\alpha$\,H$\beta$2}}$        & 0.041 &$\pm$& 0.010 \\
R$_{\text{H$\alpha$\,H$\gamma$}}$        & 0.013 &$\pm$& 0.003 \\
R$_{\text{H$\beta$2\,H$\gamma$}}$        & 0.040 &$\pm$& 0.020 \\
R$_{\text{H$\alpha$\,H$\delta$1}}$       & 0.0   &$+$&   0.005 \\
R$_{\text{H$\alpha$\,H$\delta$2}}$       & 0.014 &$\pm$& 0.003 \\
R$_{\text{H$\beta$1\,H$\delta$1}}$       & 0.023 &$\pm$& 0.010 \\
R$_{\text{H$\beta$1\,H$\delta$2}}$       & 0.0   &$+$&   0.002 \\
R$_{\text{H$\beta$2\,H$\delta$1}}$       & 0.0   &$+$&   0.002 \\
R$_{\text{H$\beta$2\,H$\delta$2}}$       & 0.022 &$\pm$& 0.010 \\
$^3$J$_{\text{H$\alpha$\,H$\beta$1}}$    & 7.381 &$\pm$& 0.042 \\
$^3$J$_{\text{H$\alpha$\,H$\beta$2}}$    & 4.549 &$\pm$& 0.034 \\
$^3$J$_{\text{H$\beta$1\,H$\gamma$}}$    & 4.999 &$\pm$& 0.034 \\
$^3$J$_{\text{H$\beta$2\,H$\gamma$}}$    & 8.070 &$\pm$& 0.059 \\
$^3$J$_{\text{H$\alpha$\,C$\gamma$}}$    & 3.0   &$\pm$& 2.0   \\
$^3$J$_{\text{C$\alpha$\,H$\gamma$}}$    & 5.0   &$\pm$& 2.0   \\
$^3$J$_{\text{C$^{\prime}$\,H$\beta$1}}$ & 5.3   &$\pm$& 0.5   \\
$^3$J$_{\text{C$^{\prime}$\,H$\beta$2}}$ & 4.2   &$\pm$& 0.5   \\
\end{tabular}
\end{ruledtabular}
\end{table}

The experimental \jj{H$\alpha$}{H$\beta$} and \jj{H$\beta$}{H$\gamma$}
vicinal coupling constants (Table \ref{tab2:NMRdata})
show the doubly conjugate pattern
expected for leucine side chains, that is strong -- weak and weak -- strong.
The approximately 3 Hz difference between the weak and strong
couplings indicates that both the trans gauche$^+$ and gauche$^-$ trans
leucine side chain isomers are significantly populated and
that the $\beta$-proton assignment is best determined by comparing
the goodness-of-fit of the two alternative assignments.
For the preliminary $\beta$-proton assignment in this section
it is adequate to fit only the \jj{H$\alpha$}{H$\beta$}
and \jj{C$^{\prime}$}{H$\beta$} coupling constants.
The resulting two rotational isomer model, see methods,
does not have any dependence on the $\chi^2$ torsion angle; nevertheless,
throughout this paragraph we maintain the assumption of highly correlated
$\chi^1$ and $\chi^2$ torsion angles and continue to refer to these
two rotational isomers as trans gauche$^+$ and gauche$^-$ trans.
The goodness-of-fit of the simple two rotational isomer
model is $2\times 10^{-2}$ for the $\beta$-proton assignment
in Table \ref{tab1:assignments} and $2\times 10^{-3}$
for the alternative assignment.
The better fit gives population estimates of 39\% trans gauche$^+$
and 61\% gauche$^-$ trans with an uncertainty of $\pm$9\%.
These population estimates fall in the gray area between
predominantly gauche$^-$ trans and approximately
equal mixture of both conformations.
On either side of this gray area the assignment made by
inspection agrees with that obtained by fitting the experimental
\jj{H$\alpha$}{H$\beta$} and \jj{C$^{\prime}$}{H$\beta$} coupling constants.
Suppose gauche$^-$ trans predominates.
Then the $\chi^1$ torsion angle is gauche$^-$,
H$^{\alpha}$ and H$^{\beta 1}$ are antiperiplanar,
the carboxyl carbon and both $\beta$-protons are synclinal,
and the assignment in Table \ref{tab1:assignments} is correct because
the \jj{H$\alpha$}{H$\beta$1} coupling is stronger than
the \jj{H$\alpha$}{H$\beta$2} coupling and
both \jj{C$^{\prime}$}{H$\beta$} couplings are fairly weak.
On the other hand suppose the two conformations are approximately
equally mixed.
Then the carboxyl carbon and the $\beta$1-proton are synclinal
in one conformation and antiperiplanar in the other,
but the carboxyl carbon and the $\beta$2-proton are synclinal
in both conformations,
and the assignment in Table \ref{tab1:assignments} is again correct
because the \jj{C$^{\prime}$}{H$\beta$1} coupling is stronger
than the \jj{C$^{\prime}$}{H$\beta$2} coupling.

The goodness-of-fit of the simple two rotational isomer model
is only $2\times 10^{-2}$ because this model predicts 
a high average \jj{H$\alpha$}{H$\beta$} coupling constant
and too low a \jj{C$^{\prime}$}{H$\beta$2} coupling constant.
As noted above the average of the two \jj{H$\alpha$}{H$\beta$}
coupling constants is $5A/8+B/4+C$ for ideal geometry.
Karplus coefficients for this coupling \cite{Cung82,DeMarco78,Kessler87}
give average values ranging from 8.1 to 8.7 Hz.
These predicted values must be compared with 6.0 Hz, which is the average
of the two experimental \jj{H$\alpha$}{H$\beta$}
couplings in Table \ref{tab2:NMRdata}.
One explanation for this difference is that there is a small population
of rotational isomers with gauche$^+$ $\chi^1$ torsion angle.
This would reduce the predicted average coupling constant
because both $\beta$-protons are synclinal to the $\alpha$-proton
when $\chi^1$ is gauche$^+$.
What is important is not the magnitude in Hertz of the difference
between the average predicted and experimental \jj{H$\alpha$}{H$\beta$}
couplings, but the standard deviation, that is, the ratio of this difference
over the estimated error.
The one and one-half standard deviation difference found here
is not improbably large and reflects our estimate
of the errors in the Karplus equation calibration, see background section,
and of the errors due to the assumption of ideal geometry.
In view of the known improbablity of leucine gauche$^+$ $\chi^1$
rotational isomers, again see background section,
these last two sources of error are a more likely
explanation of the difference.

The above explanation of the difference between the
average predicted and experimental \jj{H$\alpha$}{H$\beta$}
coupling constants is even more plausible
in view of a similar difference found
between the average predicted and experimental
\jj{H$\beta$}{H$\gamma$} coupling constants.
Though the dipeptide backbone conformation leaves some room for doubt about
the complete absence of $\chi^1$ is gauche$^+$ conformation,
the evidence from crystallographic studies and conformational
analysis is very good that there is at most a very small population
of rotational isomers with a gauche$^-$ $\chi^2$ torsion angle.
Also the cobalt complex with the dipeptide backbone should have relatively
little effect on the \jj{H$\beta$}{H$\gamma$} coupling constant.
For ideal geometry the average of the two
\jj{H$\beta$}{H$\gamma$} coupling constants is given by the
same expression as for the \jj{H$\alpha$}{H$\beta$} average.
Karplus coefficients for the {\em sec}-butyl coupling \cite{BothnerBy65}
give an average value of 8.5 Hz and coefficients corrected
for substituent electronegativity as suggested by Pachler
(Ref.\ \onlinecite{Pachler72}, Eq.\ 2 and Table 4)
give an average value of 8.1 Hz.
The average of the two experimental couplings
in Table \ref{tab2:NMRdata} is 6.5 Hz.
Error in the Karplus coefficient calibration
and perhaps some departure from ideal geometry seem to be the only plausible
explanation is this difference.  This supports our view that
overall errors of one to two Hertz are entirely possible.
The two rotational isomer model also
predicts that the \jj{C$^{\prime}$}{H$\beta$2}
coupling constant is $A/4-B/2+C$ because the $\beta$2-proton
is always synclinal to the leucine carboxyl carbon when
rotational isomers with a gauche$^+$ $\chi^1$ torsion angle are excluded.
The predicted coupling \cite{Fischman80} is 1.4 Hz and the observed
is 4.2 Hz (Table \ref{tab2:NMRdata}).  A small gauche$^+$ $\chi^1$ population
could make up much of this two standard deviation difference,
but we again favor the explanation that the Karplus calibration
is not very accurate.

The $\delta$-proton assignments in Table \ref{tab1:assignments}
follow from the pattern of R$_{\text{H$\beta$H$\delta$}}$ cross
relaxation rates in Table \ref{tab2:NMRdata}.
These assignments are also model dependent.
For several models the $\delta$-proton assignment is unambiguous
once the $\beta$-proton assignment is selected (results not presented);
however, to keep things simple we again assume
that only the trans gauche$^+$ and gauche$^-$ trans
leucine side chain rotational isomers are significantly populated.
For both these rotational isomers the H$^{\beta 1}$ to H$^{\delta 1}$
and H$^{\beta 2}$ to H$^{\delta 2}$ distances are 2.8 to 2.9 angstroms.
The H$^{\beta 1}$ to H$^{\delta 2}$ and H$^{\beta 2}$ to H$^{\delta 1}$
distances are 2.9 and 4.0 angstroms for the trans gauche$^+$ rotational isomer
and reverse order to 4.0 and 2.9 angstroms for the gauche$^-$ trans isomer.
The $\delta$-proton assignments in Table \ref{tab1:assignments} produce the
strong -- weak -- weak -- strong pattern observed in the experimental
relaxation rates in Table \ref{tab2:NMRdata}.

An unambiguous $\beta$-proton assignment is not possible
if an arbitrarily large population of gauche$^+$ $\chi^1$
rotational isomers is allowed.
We have repeated the above least squares fit to the
experimental \jj{H$\alpha$}{H$\beta$} and \jj{C$^{\prime}$}{H$\beta$2}
coupling constants, while allowing rotational isomers with
gauche$^-$, gauche$^+$, and trans $\chi^1$ torsion angles.
For the $\beta$-proton assignment in Table \ref{tab1:assignments}
the goodness-of-fit of this
three rotational isomer model, see methods, rises to 29\%
and that of the alternative $\beta$-proton assignment rises
to 94\%.  By this criterion either assignment is now acceptable.
The population of rotational isomers with gauche$^+$ $\chi^1$ torsion angles
is 36\% for the selected assignment and 51\% for the alternative.
Either of these gauche$^+$ populations seems unacceptably high.
In any event the presently available models are not able to meaningfully
predict the gauche$^+$ $\chi^1$ population.
The experimental data can be satisfactorily explained
by a two rotational isomer model,
which excludes gauche$^+$ $\chi^1$ rotational isomers.
The unambiguous assignment of the $\beta$ and $\delta$-protons 
probably must await the preparation of cobalt dipeptide with
stereoselectively deuterated leucine side chains \cite{Ostler93}.

\section{COMPUTATIONAL RESULTS AND DISCUSSION}

\subsection{Chelate ring conformation}

As discussed in the background section, conformational analysis
predicts that gauche$^+$ $\chi^1$ rotational isomers
of the leucine side chain are favored in the absence
of chelate ring puckering.
Crystallographic and NMR evidence shows that
the cobalt dipeptide chelate rings do pucker and that the
$\phi_2$ and $\psi_2$ torsion angles depart from 180 degrees
by as much as 10 or 20 degrees.
Simple inspection of backbone-dependent rotamer libraries
suggests that this departure is large enough
to reduce the population of gauche$^+$ $\chi^1$ rotational isomers
to fairly low levels.
To obtain a sharper picture of the dependence on backbone conformation
of rotamer preferences we have constructed a rotamer
library for a region-of-interest around the special point
in $\phi \times \psi$ space where gauche$^+$ $\chi^1$
rotamers are most favored, that is, the point with coordinates
$\phi = -175$ and $\psi = 175$ degrees.
This region-of-interest rotamer library differs from previous
backbone-dependent rotamer libraries \cite{Dunbrack94,Dunbrack93}
only in that a limited region of $\phi \times \psi$ space
is divided into annular disks around the special point
instead of dividing the entire $\phi \times \psi$ space
into a grid of square cells.
Our region-of-interest rotamer library is constructed from a list
of backbone and side chain torsion angles of 7085 leucine residues
on 445 nonhomologous (that is with less than 50\% sequence identity)
protein chains from a recent Brookhaven Protein Database of structures
with a resolution of 2.0 angstroms or better.
Backbone angles in the region-of-interest are
not very common in protein structures.
There are only 0, 13, 28, 72, 112, and 251 of the leucine
residues with backbone $\phi$ and $\psi$ angles in the 6 annular shells
with a width of 10 degrees and that have outer radii
of 10, 20, 30, 40, 50, and 60 degrees.
There are 11 residues with a gauche$^+$ $\chi^1$ torsion angle out of
the 13 (92\%) with backbone torsion angle in the 10 to 20 degree annulus,
21 of 28 (75\%) in the 20 to 30 annulus,
29 of 72 (40\%) in the 30 to 40 annulus,
10 of 112 (9\%) in the 40 to 50 annulus, and
7 of 251 (3\%) in the 50 to 60 annulus.
The region-of-interest rotamer library shows a dramatic drop-off
of gauche$^+$ $\chi^1$ leucine side chain rotational isomers when the backbone
torsion angle is beyond the 20 to 30 degree annulus.
Note that $\chi^2$ torsion angles of most of the gauche$^+$ $\chi^1$
rotational isomers in this region-of-interest are also gauche$^+$.

\begin{figure}
\resizebox{8.3cm}{!}{\includegraphics{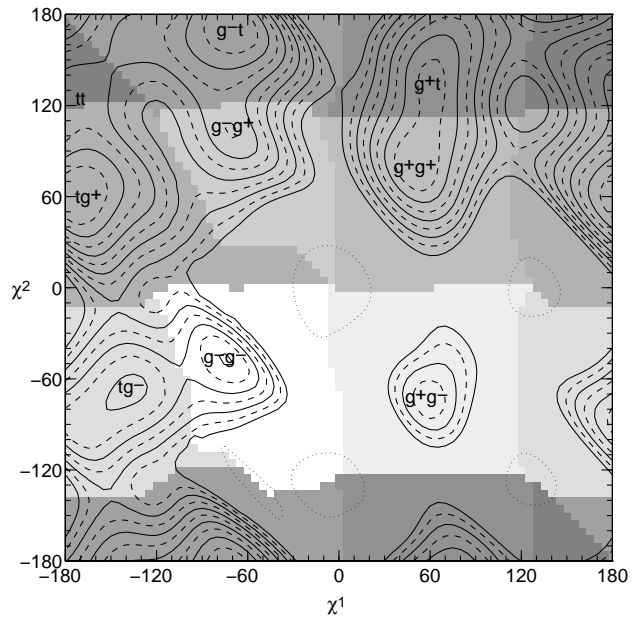}}
\caption{Molecular mechanics energy map for rotational
isomerization of cobalt dipeptide leucine side chain.
Contour levels are {\em dashed}, 1, 3, 5, 7, 9;
{\em solid}, 2, 4, 6, 8, 10 kcal/mol.  Zero corresponds to $-$39.4 kcal/mol.
The nine rotational isomers are labeled at the position
of their energy well minima.
The background shading shows the energy well boundaries
and the torsion space regions for averaging the NOESY cross relaxation
and vicinal coupling constants.}
\label{fig2:contour_plot_x1x2}
\end{figure}

The conformational statistics of leucines in protein database structures
show clearly that side chain rotamer preferences are highly sensitive
to the backbone conformation, especially near the $\phi$ and $\psi$ angles
of the cobalt dipeptide backbone.
Analysis of the molecular mechanics energy map over $\chi^1 \times \chi^2$
torsion space of the leucine side chain (Fig.\ \ref{fig2:contour_plot_x1x2})
and of the backbone conformation of the energy minimized
dipeptide structures suggests that
the cobalt dipeptide chelate rings do indeed pucker enough
to allow gauche$^-$ and trans $\chi^1$ rotational isomers to predominate.
The gauche$^-$ trans energy well has the lowest energy minimum,
which we assign the value exactly 0 kcal/mol, and the trans gauche$^+$
well is only 0.1 kcal/mol higher.
The energies of the three energy well minima with the
$\chi^1$ torsion angle gauche$^+$ are
6.4 kcal/mol for gauche$^+$ gauche$^-$,
2.1 kcal/mol for gauche$^+$ gauche$^+$, and
2.6 kcal/mol for gauche$^+$ trans.
All other well minima have energies higher than 2.9 kcal/mole.
The molecular mechanics energy map prediction matches the
backbone-independent conformational analysis result
that the trans gauche$^+$ and gauche$^-$ trans
leucine side chain rotational isomers are the most stable.
The backbone torsion angles of the minimized structures
at the energy minimum grid point of the gauche$^-$ trans energy well
are $\phi = -162$ and $\psi = 167$ degrees and
of the trans gauche$^+$ well are $\phi = -166$ and $\psi = 168$ degrees.
These leucine backbone torsion angles specify a point
in $\phi \times \psi$ space that is 15 and 11 degrees
from the point where gauche$^+$ $\chi^1$
rotamers are most favored, that is, the point with coordinates
$\phi = -175$ and $\psi = 175$ degrees.
The backbone torsion angles at the energy minimum grid point
of the three gauche$^+$ $\chi^1$ rotational isomers
are all in the ranges $-176 \le \phi \le -174$ and $178 \le \psi \le 180$
and are all within 4 to 6 degrees of the
point with both $\phi$ and $\psi = 180$ degrees.
This seems to confirm that the backbone torsion angles
of the trans gauche$^+$ and gauche$^-$ trans rotational isomers
are indeed adjustments from cobalt dipeptide chelate ring planar geometry
that accommodate unfavorable backbone-dependent {\em syn}-pentane interactions.
To eliminate these backbone-dependent interactions
the backbone conformation apparently need adjust only
by an angle of 10 to 15 degrees in $\phi \times \psi$
torsion space, which is about half that suggested
by the region-of-interest rotamer library.

To definitively establish the amount of chelate ring puckering
and its influence on leucine side chain populations
will require more extensive molecular mechanics calculations
and more reliable energy map error estimates than presented here.
Such molecular mechanics studies are important because
as we have already emphasized in the experimental section
the NMR data by itself does not give an unambiguous assignment
of the $\beta$ and $\delta$-protons.
Further molecular mechanics studies are also needed to validate
our analysis on measurabality of rotational isomer populations,
which is presented in the concluding subsection of this results
and discussion section.
There we analyze the measurability of the gauche$^+$ gauche$^+$
rotational isomer population based on the assumption that the ratio
of gauche$^+$ to trans or gauche$^-$ $\chi^1$ rotational
isomer populations is small.  The assignments presented
in the experimental section also rely on this assumption.

The accuracy of molecular mechanics predictions of the relative
populations of $\chi^1$ rotational isomers
depends on achieving the correct balance of at least three energy terms:
the steric energy of {\em syn}-pentane interactions,
the energy of 10 to 15 degree compensatory rotations
of the leucine $\chi^1$ and $\chi^2$ side chain dihedral angles,
and the energy of ring puckering associated with the rotation
of the leucine $\phi$ and $\psi$ backbone dihedral angles.
Two general considerations suggest that these three energy terms
are correctly balanced in the present molecular mechanics calculations.
First, {\em syn}-pentane interaction accounting correctly predicts
the observed rotamer preferences of protein side chains \cite{Dunbrack94}.
This implies that compensatory rotations of the side chain
dihedral angles don't significantly deminish the importance
of the {\em syn}-pentane effect in predicting rotamer preferences
and that the balance of the first two of the three above energy
terms is at least qualitatively correct in our calculations.
Second, energy minimization of pentane structures with CHARMM
parameters \cite{MacKerell98}, which are very similar to those
we employ, quantitatively reproduces the conformational
energies observed experimentally or predicted by {\em ab initio}
calculations \cite{Dunbrack94}.
This suggests that the balance of the first two of the three above energy
terms is also quantitatively correct in our calculations.
It remains to establish the accuracy of the last of the above
three energy terms, the energy of ring puckering associated with
the rotation of the leucine $\phi$ and $\psi$ backbone dihedral angles.
The molecular mechanics parameters of the cobalt chelate ring
complex are expected to play and important role in determining
the ring puckering energy.
Most of the bond length and bond angle molecular mechanics parameters
are known from previous crystallographic \cite{Prelesnik84}
or molecular mechanics \cite{Buckingham74} studies of cobalt complexes.
At the other extreme the torsion angle and improper torsion angle
force constants with cobalt in one of the four angle defining positions
and charges of the nitro and cobalt atoms are not much better than
order of magnitude guesses.
Furthermore the $-2$ charge distributed over the cobalt complex
may introduce a substantial solvation effect \cite{Schaefer98,Gresh98}
into the ring puckering energy.  Indeed the solvation effects
may be viewed as a fourth energy term that affects the accuracy
of molecular mechanics predictions of the relative
populations of $\chi^1$ rotational isomers.

More extensive molecular mechanics studies are certainly
needed to establish the effect molecular mechanics
parameters and solvation have on the relative isomer populations.
However, some simple tests of the
present molecular mechanics suggest we aren't too far off.
The molecular mechanics energy map over $\chi^1 \times \chi^2$
torsion space of the leucine side chain is not very sensitive
to the values of the uncertain parameters.
The relative energy well depths
of the leucine side chain rotational isomers
vary by less than about one half kcal/mol
when the torsion angle and improper torsion angle
force constants involving cobalt are scaled down to zero as a group
or when the distance independent dielectric constant equal to one
is replaced by a distance dependent dielectric constant equal
to the inverse atomic separation in angstroms.

\subsection{Effect of intramolecular motions}

For most leucine side chain rotational isomers the effect
of thermal motions on the NOESY cross relaxation rates
is several times smaller than the typical accuracy of these measurements
and the effect on vicinal coupling constants is perhaps several
times bigger than the accuracy of the best homonuclear coupling measurements.
The effect of thermal motions on side chain vicinal coupling
constants is similar in magnitude to the previously reported
effect on backbone coupling constants \cite{Wang96}.
The magnitude of the thermal motion effect is estimated by comparing
the calculated average NMR observables to those values calculated
at the average $\chi^1$ and $\chi^2$ torsion angles.
These averages are taken over the individual energy well regions,
see methods, and thus the comparison looks at the effects of
fast thermal motions within the energy wells as opposed to the
effects of slower interconversion of rotational isomers.
The trans gauche$^+$ and gauche$^-$ trans rotational isomers,
which are the predominantly populated isomers, are typical.
For these two rotational isomers the RMS differences
between the average observables and the observables
at the average are 0.16 Hz for the vicinal coupling constants
and 0.0011 s$^{-1}$ for the NOESY cross relaxation rates,
where these RMS differences are averaged over these two rotational
isomers and over the 10 NOESY cross relaxation rates
and 8 vicinal coupling constants listed in Table \ref{tab2:NMRdata}.
These differences are substantially increased for rotational
isomers with more anharmonic $\chi^1 \times \chi^2$
torsion space energy wells.
The difference between the energy well minimum position
and the average $\chi^1$ and $\chi^2$ torsion angles gives a rough measure
of the anharmonicity of a rotational isomer energy well.
By this measure the trans gauche$^-$ energy well is the most anharmonic
(compare Fig.\ \ref{fig2:contour_plot_x1x2}) with a difference of 12 degrees
between the minimum and average positions.
For the trans gauche$^+$ and gauche$^-$ trans energy wells
these differences are only 3 and 4 degrees.
For the anharmonic energy well of the trans gauche$^-$ rotational isomer
the RMS differences between the average observables and the observables
at the average are 0.31 Hz for the vicinal coupling constants
and 0.0021 s$^{-1}$ for the NOESY cross relaxation rates,
which are both almost twice the values for the more nearly
harmonic energy wells of the trans gauche$^+$ and gauche$^-$ trans
rotational isomers.

The difference between the average observables and the observables
at the average $\chi^1$ and $\chi^2$ torsion angles only accounts
for about half the effect of thermal motions on the NMR observables.
Thermal shifting of the average $\chi^1$ and $\chi^2$ torsion angles
of a rotational isomer also significantly changes the NMR observables.
This thermal motion effect is measured by the difference
between the NMR observables at the $\chi^1 \times \chi^2$ energy well
minimum and at the average $\chi^1$ and $\chi^2$ torsion angles.
As already noted in the last paragraph the difference
between the minimum and average positions of the
trans gauche$^+$ and gauche$^-$ trans energy wells
is 3 and 4 degrees.  For these rotational isomers the
RMS differences between the observables at the minimum and average
positions are 0.17 Hz for the vicinal coupling constants and
0.0013 s$^{-1}$ for the NOESY cross relaxation rates.
These differences are about the same as the values
of 0.16 Hz and 0.0011 s$^{-1}$ reported in the previous paragraph
for the differences between the average observables
and the observables at the average.
In the case of the anharmonic energy well of the trans gauche$^-$
rotational isomer the thermal shifting effect is more dramatic.
The RMS differences between the observables at the minimum and average
positions are 0.75 Hz and 0.0041 s$^{-1}$, which are about twice
the differences between the average observables
and the observables at the average for the trans gauche$^-$ rotational isomer.

All these thermal motion effects can be put into perspective
by comparing them to the differences in the NMR observables
at the $\chi^1 \times \chi^2$ energy well minimum
and at ideal geometry $\chi^1$ and $\chi^2$ torsion angles.
For the trans gauche$^+$ and gauche$^-$ trans rotational isomers
the RMS differences between the minimum energy and ideal geometry
observables are 0.50 Hz for the vicinal coupling constants
and 0.0039 s$^{-1}$ for the NOESY cross relaxation rates,
which is about three times the size of the thermal motion effect.
This is just about what might be expected because the
energy minima of the trans gauche$^+$ and gauche$^-$ trans rotational
isomers differ by 9 and 12 degrees from the ideal geometry
positions in $\chi^1 \times \chi^2$ torsion space and these
differences are three times the differences between
the average $\chi^1$ and $\chi^2$ torsion angles and
the positions of the energy well minima.

\subsection{Necessity of molecular mechanics energy estimates}

\begin{table*}
\caption{\label{tab3:ModelsSummary}Rotational isomer populations.
Each row shows the fit populations for a model that excludes the
dotted rotational isomers.
All population estimates are in percent with the error in the last
digit given in parenthesis.}
\begin{ruledtabular}
\begin{tabular}{r@{\extracolsep{24pt}}*{9}{r@{\extracolsep{1pt}}l@{\extracolsep{12pt}}}d@{\extracolsep{6pt}}lll}
   &\multicolumn{2}{l}{ g$^-$g$^-$ }&\multicolumn{2}{l}{ g$^+$g$^-$ }&\multicolumn{2}{l}{\mbox{$\;$tg$^-$}}&
    \multicolumn{2}{l}{ g$^-$g$^+$ }&\multicolumn{2}{l}{ g$^+$g$^+$ }&\multicolumn{2}{l}{\mbox{$\;$tg$^+$}}&
    \multicolumn{2}{l}{ g$^-$t     }&\multicolumn{2}{l}{ g$^+$t     }&\multicolumn{2}{l}{\mbox{$\;$tt    }}&
$Q$\footnotemark[1]&\multicolumn{3}{c}{ } \\ \hline
 1& 8&(8) & 0&(8) & 5&(9) & 4&(7) & 9&(7) & 8&(8) &30&(10)&22&(12)&13&(9) &50.0&MC\footnotemark[2]& & \\
 2&\J&    &\J&    &52&    &\J&    &\J&    &\J&    &48&    &\J&    &\J&(6)\footnotemark[3]
                                                                          & 6.0&  & & \\
 3&\A&$\!$\footnotemark[4]
          &50&    &\J&    &\J&    &\J&    &\J&    &50&    &\J&    &\J&(6) & 3.1&  & & \\
 4&\J&    &\J&    &\J&    &\A&    &43&    &\J&    &57&    &\J&    &\J&(5) & 2.1&  & & \\
 5&\J&    &\J&    &\J&    &\J&    &\J&    &37&    &63&    &\J&    &\J&(4) & 2.0\footnotemark[5]&  & & \\
 6&\J&    &\A&    &\J&    &39&    &\A&    &\J&    &\J&    &61&    &\J&(6) & 1.3&  & & \\
 7&\J&    &\J&    &\J&    &\J&    &\J&    &25&    &39&    &35&    &\J&(7) &68.7&  & & \\
 8&\J&    &\J&    &\J&    &\A&    &29&    &\A&    &46&    &\J&    &26&(6) &67.8&  & & \\
 9&\J&    &33&    &\J&    &\J&    &\J&    &\J&    &42&    &\J&    &25&(6) &64.1&  & & \\
10&\A&    &30&    &\J&    &\A&    &\A&    &22&    &48&    &\J&    &\J&(6) &52.7&  & & \\
11&\A&    &\J&    &32&    &\J&    &23&    &\J&    &44&    &\J&    &\J&(7) &42.8&  & & \\
12&\J&    &\J&    &36&    &\J&    &\J&    &\J&    &37&    &28&    &\J&(8) &26.1&  & & \\
13&\J&    &\J&    &\J&    &\J&    &\J&    &\J&    &38&    &35&    &27&(7) &25.2&  & & \\
14&36&    &\A&    &\J&    &\J&    &\J&    &\J&    &\J&    &38&    &26&(7) &24.7&  & & \\
15&\A&    &26&    &31&    &\J&    &\J&    &\J&    &44&    &\J&    &\J&(9) &22.2&  & & \\
16&34&    &\A&    &\J&    &\J&    &\A&    &21&    &\J&    &45&    &\J&(7) &21.3&  & & \\
17&\J&    &\A&    &30&    &33&    &\A&    &\J&    &\J&    &37&    &\J&(8) &20.5&  & & \\
18&\A&    &\J&    &46&    &20&    &\J&    &\J&    &34&    &\J&    &\J&(8) &18.1&  & & \\
19&\J&    &\J&    &\J&    &\A&    &24&    &21&    &54&    &\J&    &\J&(7) &17.7\footnotemark[6]&  & & \\
20&\J&    &\J&    &32&    &\J&    &\J&    &17&    &50&    &\J&    &\J&(8) &13.8&  & & \\
21&\J&    &\A&    &\J&    &33&    &\A&    &\J&    &\J&    &46&    &21&(7) &13.4&  & & \\
22&\J&    &\J&    &48&    &\J&    &\J&    &\J&    &52&    &\J&    &\J&(6) &11.1&  & &A\footnotemark[7]\\
23&\J&    &\J&    &\J&    &\J&    &\J&    &37&    &63&    &\J&    &\J&(4) & 6.3\footnotemark[8]
                                                                               &  & &A\\
24&\J&    &\J&    &\J&    &\A&    &44&    &\J&    &56&    &\J&    &\J&(5) & 4.4&  & &A\\
25&\A&    &49&    &\J&    &\J&    &\J&    &\J&    &51&    &\J&    &\J&(6) & 3.5&  & &A\\
26&\J&    &\A&    &\J&    &40&    &\A&    &\J&    &\J&    &60&    &\J&(6) & 1.6&  & &A\\
27&\J&    &\J&    &\J&    &\J&    &\J&    &46&    &\J&    &54&    &\A&(5) &65.2&  &R\footnotemark[9]& \\
28&\J&    &\J&    &\J&    &\J&    &59&    &\J&    &\J&    &\J&    &41&(5) &29.9&  &R& \\
29&\J&    &57&    &\J&    &\J&    &\J&    &43&    &\J&    &\J&    &\J&(6) &17.0&  &R& \\
30&\J&    &\A&    &52&    &\J&    &48&    &\J&    &\J&    &\J&    &\J&(7) &15.3&  &R& \\
31&\J&    &\J&    &\J&    &\J&    &\J&    &37&(4) &63&(4) &\J&    &\J&    & 2.0&MC& & \\
32&\J&    &\J&    &\J&    &\J&    &\J&    &60&(4) &40&(4) &\J&    &\J&    & 0.2&MC&R& \\
33&\J&    &\J&    &\J&    &\J&    &\J&    &37&(4) &63&(4) &\J&    &\J&    & 6.3&MC& &A\\
34&\J&    &\J&    &\J&    &\J&    &\J&    &61&(4) &39&(4) &\J&    &\J&    & 1.0&MC&R&A\\
35&\J&    &\J&    &\J&    &\J&    &24&(8) &21&(7) &54&(5) &\J&    &\J&    &17.7&MC& & \\
36&\J&    &\J&    &\J&    &\J&    &39&(8) &34&(7) &27&(5) &\J&    &\J&    &62.4&MC&R& \\
37&\J&    &\J&    &\J&    &\J&    & 3&(5) &35&(5) &61&(5) &\J&    &\J&    &\cdots\footnotemark[10]\!\!\!\!&MC& & \\
38& 0&(0) &\J&    & 0&(0) & 0&(0) & 0&(0) &37&(0) &62&(1) & 0&(0) & 0&(0) &\cdots&MC& & \\
\end{tabular}
\end{ruledtabular}
\footnotetext[1]{\begin{minipage}[t]{6.5in}\raggedright
The goodness-of-fit $Q$ is the percentage probability
that chi-square exceeds its fit value.\vspace{2pt} \end{minipage}}
\footnotetext[2]{\begin{minipage}[t]{6.5in}\raggedright
For models marked MC the population errors are calculated
from a Monte Carlo simulation of the NMR observables.\vspace{2pt} \end{minipage}}
\footnotetext[3]{\begin{minipage}[t]{6.5in}\raggedright
The population errors of models without Monte Carlo
error analysis are derived from a least squares moment matrix.
Only the average population error is reported in the last isomer column.\vspace{2pt} \end{minipage}}
\footnotetext[4]{\begin{minipage}[t]{6.5in}\raggedright
If a rotational isomer marked with an uparrow
is added to the model in this row, then a nonnegativity constraint
will fix the rotational isomer population at zero.
Because these constraints are NMR observable dependent, they are only
reported for models without Monte Carlo simulated NMR observables.\vspace{2pt} \end{minipage}}
\footnotetext[5]{\begin{minipage}[t]{6.5in}\raggedright
The (tg$^+$,g$^-$t) model is among the 5 two rotational
isomer models with a goodness-of-fit better than 1\%.\vspace{2pt} \end{minipage}}
\footnotetext[6]{\begin{minipage}[t]{6.5in}\raggedright
The (g$^+$g$^+$,tg$^+$,g$^-$t) model is among the 15 three
rotational isomer models with a goodness-of-fit better than 10\%.\vspace{2pt} \end{minipage}}
\footnotetext[7]{\begin{minipage}[t]{6.5in}\raggedright
For models marked A the NMR observables are calculated
by Boltzmann weighted averaging over the energy well of each rotational isomer.
The NMR observables of all other models are calculated at the energy
well minimum of each rotational isomer.\vspace{2pt} \end{minipage}}
\footnotetext[8]{\begin{minipage}[t]{6.5in}\raggedright
The (tg$^+$,g$^-$t) model appears again with
improved goodness-of-fit among these 5 two rotational isomer models
with a goodness-of-fit better than 1\%.\vspace{2pt} \end{minipage}}
\footnotetext[9]{\begin{minipage}[t]{6.5in}\raggedright
For models marked R the assignments of both $\beta$
and $\delta$-protons are reversed.\vspace{2pt} \end{minipage}}
\footnotetext[10]{\begin{minipage}[t]{6.5in}\raggedright
The ellipsis marks indicate that the g$^+$g$^+$
measurability simulations do not generate any additional measure
of goodness-of-fit because they are only indirectly based on the
experimental NMR observables.\vspace{2pt} \end{minipage}}
\end{table*}

At the present accuracy of Karplus equation calibrations
it is not possible to calculate the populations of all 9 rotational
isomers of the leucine side chain from only the
NMR data in Table \ref{tab2:NMRdata}.
Before fitting the NMR data a small set of rotational isomers with
nonzero population must be selected with molecular mechanics calculations
or conformational analysis.
If we calculate the population of all 9 rotational isomers
by fitting an 8 parameter model (Table \ref{tab3:ModelsSummary} row 1),
then the populations estimates range from 0.0 to 0.3,
though most of them are near 0.1, and
the population error estimates from the moment matrix
range from about $\pm$0.2 to $\pm$0.3.
The standard deviations of the Monte Carlo probability density functions
are all close to $\pm$0.1.  These errors are somewhat smaller
than suggested by the moment matrix because the moment matrix
estimates do not take into account the nonnegativity constraints
on the isomer populations.
The errors given by either set of error estimates are larger than
or at least nearly as large as the population estimates.
This indicates that fitting the 9 rotational isomer model
gives meaningless population estimates.

There are $2^9-1=511$ possible nonempty subsets
of the set of 9 rotational isomers.
As we will detail shortly, these rotational isomer subsets
generate a large number of distinct solutions to the problem
of fitting the experimental NMR data.
As an alternative to fitting all the rotational isomer populations,
we might hope to find among this large set of solutions
one best solution that includes only a small number of rotational isomers
and that has a uniquely high goodness-of-fit to the NMR data.
A solution is initially generated from each subset
of rotational isomers by fitting the experimental NMR data with
the model that includes only the isomers in that subset.
In general the populations of these included isomers
are not all positive because active nonnegativity constraints
force some populations to be exactly zero.
A different set of experimental data would yield a different
set of active constraints and thus a different set of
positive populations.
In the next subsection we fit multiple Monte Carlo
simulated NMR data sets to calculate population probability distributions.
But here we fit only the actual experimental data
and generate one single solution from each subset of rotational isomers.
Two different subsets of isomers may generate the same
solution with the same positive isomer populations
on some common subset of the two original subsets of isomers.
Thus the number of unique solutions is substantially smaller than the number
of subsets of 9 rotational isomers.
A single unique solution is conveniently identified by the positive
isomer populations and the isomers with positive populations
are referred to as the populated isomers.

For the assignments in Table \ref{tab1:assignments},
experimental data in Table \ref{tab2:NMRdata},
and with the NMR observables calculated at the $\chi^1 \times \chi^2$
energy map minimum positions the 511 isomer subsets generate
278 unique solutions.
One half of these solutions have a goodness-of-fit better than
10\% and two thirds of them have a goodness-of-fit better than 1\%.
There are 5 solutions that have only two populated isomers and have
a goodness-of-fit between 10\% and 1\% and 15 solutions that have
three populated isomers and have a goodness-of-fit better than 10\%
(Table \ref{tab3:ModelsSummary} rows 2 through 21).
Apparently many good solutions exist even with the restriction to solutions
that have only a small number of populated isomers.
Worse, the solutions with two or three populated isomers
are inconsistent and over-fit the experimental data.
As a group, the 5 good solutions with two populated isomers
give 5 predictions of the population of each of the 9 rotational isomers.
The only consistent predictions are for the gauche$^-$ gauche$^-$ and
trans trans rotational isomer populations, which all 5 solutions
predict are zero.
The gauche$^-$ trans rotational isomer population is predicted
4 times in the range 0.5 to 0.6 and once at zero.
The other 6 rotational isomer populations are each predicted
once in the range 0.4 to 0.6 and 4 times at zero. 
These positive and zero population predictions
are inconsistent because the positive population
errors of the 5 solutions are all around $\pm$0.05.
The 15 good solutions with three populated isomers give a similar picture.
The gauche$^-$ trans rotational isomer population is predicted
11 times in the range 0.3 to 0.5 and 4 times at zero.
The other 8 rotational isomer populations are each predicted
in the range of 0.3 or 0.4 from 2 to 7 times and otherwise at zero.
Again these population predictions are inconsistent because
the positive population errors of the 15 solutions
are all around $\pm$0.08.
The experimental NMR data is over-fit in the sense that
the predicted isomer populations depend on the model
and the discrepancies between these predictions are much larger
than the errors estimated from the fit of a single model.

The number of solutions with two or three populated isomers,
the goodness-of-fits, population predictions, and error estimates
reported in the last paragraph change very little
if the average NMR observables are fit instead of those
at the energy map minimum positions, compare
rows 2 through 6 and 22 through 26 of Table \ref{tab3:ModelsSummary}.
Inconsistent and over-fit solutions also result
if the $\beta$ and $\delta$-proton assignments
in Table \ref{tab1:assignments} are reversed.
For example, for the reverse assignments the 511 isomer subsets
generate 285 unique solutions with an overall pattern
of goodness-of-fits similar to the assignments in Table \ref{tab1:assignments}.
There are 4 solutions that have only two populated isomers
and have a goodness-of-fit better than 10\%
(Table \ref{tab3:ModelsSummary} rows 27 through 30).
The gauche$^+$ gauche$^+$ and trans gauche$^+$ rotational isomer
populations are predicted 2 times in the range 0.4 to 0.6 and once at zero.
Four other rotational isomer populations
are each predicted once in the range 0.4 to 0.6 and 3 times at zero.
All 4 solutions predict that the populations of the remaining
3 rotational isomers are zero.
These population predictions are inconsistent because
the positive population errors of the 4 solutions
are all around $\pm$0.05.
Again there is no best solution and the solutions
that have only a small number of populated isomers are over-fit.
At present the only hope for obtaining a reasonable solution
is to narrow down the number of possible isomers
with the help of molecular mechanics energies or conformational analysis.

\subsection{Measurability of rotational isomer populations}

\begin{table}
\caption{\label{tab4:KarplusCoefficients}Karplus coefficients. The vicinal coupling
constants are given by the Karplus equation
$^3J(\phi)$ $=$ $A\cos^2\phi$ $-$ $B\cos\phi$ $+$ $C$,
where $\phi$ is the torsion angle between the atoms indicated in the
first column.  The coefficient units are Hz.
Throughout this work these coefficients are used to fit rotational
isomer populations to experimental vicinal coupling constants.}
\begin{ruledtabular}
\begin{tabular}{l@{\extracolsep{24pt}}d@{\extracolsep{16pt}}ddl}
&A&B&C& \\ \hline
\begin{tabular}{l} $^3$J$_{\text{H$\alpha$\,H$\beta$1}}$ \\
                     $^3$J$_{\text{H$\alpha$\,H$\beta$2}}$ \end{tabular} &
10.2 & 1.8 & 1.9 &
\footnote{Cung and Marraud \cite{Cung82}.} \\
\begin{tabular}{l} $^3$J$_{\text{H$\beta$1\,H$\gamma$}}$ \\
                     $^3$J$_{\text{H$\beta$2\,H$\gamma$}}$ \end{tabular} &
10 & 1 & 2 &
\footnote{Bothner-By (Ref.\ \onlinecite{BothnerBy65} p. 205),
Pachler (Ref.\ \onlinecite{Pachler72} p. 1939).} \\
\begin{tabular}{l} $^3$J$_{\text{H$\alpha$\,C$\gamma$}}$ \\
                     $^3$J$_{\text{C$\alpha$\,H$\gamma$}}$ \end{tabular} &
7.12 & 1.00 & 0.70 &
\footnote{Breitmaier and Voelter \cite{Breitmaier90},
Wasylishen and Schaefer (Ref.\ \onlinecite{Wasylishen73} p. 963 and
Ref.\ \onlinecite{Wasylishen72} p. 2711).} \\
\begin{tabular}{l} $^3$J$_{\text{C$^{\prime}$\,H$\beta$1}}$ \\
                     $^3$J$_{\text{C$^{\prime}$\,H$\beta$2}}$ \end{tabular} &
7.20 & 2.04 & 0.60 &
\footnote{Fischman et al. (Ref.\ \onlinecite{Fischman80} Table III).} \\
\end{tabular}
\end{ruledtabular}
\end{table}

An analysis of all the NOESY cross relaxation rates
and vicinal coupling constants listed in Table \ref{tab2:NMRdata}
and the Karplus coefficients listed in Table \ref{tab4:KarplusCoefficients}
confirms the preliminary analysis in the experimental results section
that the cobalt dipeptide leucine side chain predominantly
populates the trans gauche$^+$ and gauche$^-$ trans rotational isomers 
in approximately equal proportions.
The goodness-of-fit of this simple two rotational isomer model with
the NMR observables calculated at the $\chi^1 \times \chi^2$
energy map minimum positions is 0.020 for the assignments
given in Table \ref{tab1:assignments} and $2.2 \times 10^{-3}$ if both
the $\beta$ and $\delta$-proton assignments are reversed
(Table \ref{tab3:ModelsSummary} rows 31 and 32).
These goodness-of-fits increase to 0.063 and 0.010
if the average NMR observables are fit instead of those
at the energy map minimum positions
(Table \ref{tab3:ModelsSummary} rows 33 and 34).  For the assignments 
in Table \ref{tab1:assignments} the gauche$^-$ trans rotational isomer
predominates with a population of $0.625 \pm 0.043$.
Switching to average NMR observables has no effect on
this population or uncertainty, they both increase by only $0.001$.
The reverse assignment approximately reverses the populations,
but again does not change the uncertainty.

Analysis of both the protein data bank and the molecular mechanics energy map 
suggests that the third most populated rotational isomer after
trans gauche$^+$ and gauche$^-$ trans is gauche$^+$ gauche$^+$,
see the section on chelate ring conformation at the beginning
of this results and discussion section.
The gauche$^+$ gauche$^+$ rotational isomer population
is expected to be small, perhaps less than 5 or 10\%.
Because the 4.3\% population standard deviation
that is given by fitting the two rotational isomer model
is as large if not larger than the probable population,
it is unlikely that the gauche$^+$ gauche$^+$ population
can be measured by fitting the NMR data.
If a three rotational isomer model that includes gauche$^+$ gauche$^+$
is fit to the NMR data, then the gauche$^+$ gauche$^+$ population
is $0.245 \pm 0.078$ with a 0.18 goodness-of-fit
(Table \ref{tab3:ModelsSummary} row 35).
Though this gauche$^+$ gauche$^+$ population mean is high,
the population distribution is not inconsistent
with the expected small population.  The distribution
gives about a 5\% probability that the population is smaller than 10\%
and about a 1\% probability it is smaller than 5\%.
Note that these probability estimates must be taken with caution because,
as pointed out in the previous section, even models with three populated
rotational isomers over-fit the experimental data.
Indeed the high population mean seems to further suggest
that the gauche$^+$ gauche$^+$ population is poorly measured
by fitting the NMR data with the three rotational isomer model.
For the reverse assignments the gauche$^+$ gauche$^+$
population reaches the extremely implausible level of $0.385 \pm 0.078$
with a 0.62 goodness-of-fit (Table \ref{tab3:ModelsSummary} row 36).
The high goodness-of-fit for the three rotamer model with and without the
assignments reversed again suggests that the assignments
in Table \ref{tab1:assignments} must be taken with caution.

\begin{figure}
\resizebox{8.3cm}{!}{\includegraphics{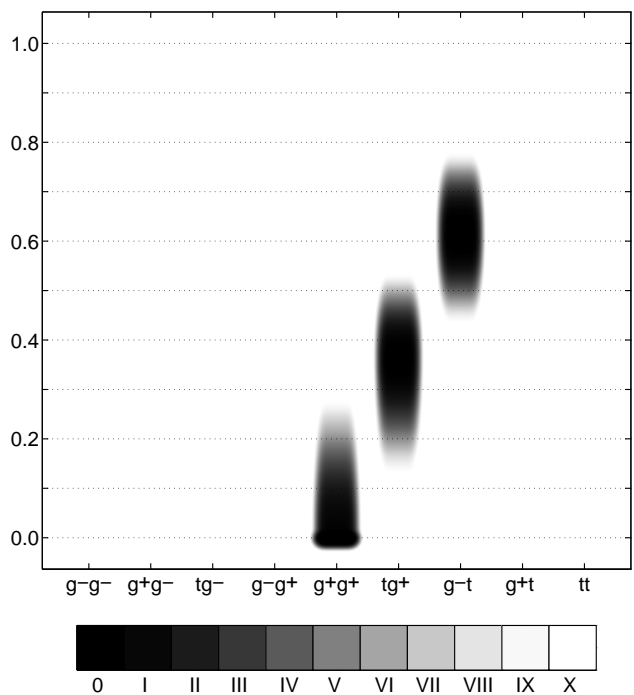}}
\caption{Gel graphic of rotational isomer population probabilities.
The graphic shows Monte Carlo probability density functions
for constrained linear least-squares parameter estimates,
where the parameters are the rotational isomer probabilities.
Each gray scale step of the stepwedge bar corresponds
to a two-fold change in probability density.
The least-squares fit has NMR measurement,
Karplus equation coefficient, and molecular mechanics geometry errors
and has the gauche$^+$ gauche$^+$ rotational isomer population fixed at zero.
The gauche$^+$ gauche$^+$ probability density function shows
the unmeasurable population range.}
\label{fig3:real_overfit}
\end{figure}

The prominent populations of the trans gauche$^+$ and gauche$^-$ trans
rotational isomers are best estimated by fitting experimental data.
The minuscule populations of the other seven rotational isomers,
except for gauche$^+$ gauche$^+$, are best estimated
from the molecular mechanics energy map.
This leaves the gauche$^+$ gauche$^+$ rotational isomer
on the awkward borderline of experimental measurability.
The standard Monte Carlo procedure \cite{Press89} can be altered to estimate
the gauche$^+$ gauche$^+$ population and its probability distribution
(Table \ref{tab3:ModelsSummary} row 37).
Initially the gauche$^+$ gauche$^+$ population is fixed at zero,
the simple two rotational isomer model is fit,
and Monte Carlo NMR observables are generated, but then these
Monte Carlo observables are fit to a three rotational isomer model
that includes gauche$^+$ gauche$^+$.
This differs from the standard procedure because one model
generates the Monte Carlo observables and a second different model
is fit to these Monte Carlo observables.  The resulting Monte Carlo
probability density functions (Fig.\ \ref{fig3:real_overfit})
give population estimates of $0.355 \pm 0.053$ and $0.612 \pm 0.045$
for the prominent trans gauche$^+$ and gauche$^-$ trans
rotational isomers and $0.033 \pm 0.046$ for the gauche$^+$ gauche$^+$
rotational isomer.
The prominent rotational isomer population estimates
have slightly smaller means and larger standard deviations
than the estimates given in the last paragraph for
the fit of the simple two rotational isomer model.
The Monte Carlo probability density function
of the gauche$^+$ gauche$^+$ rotational isomer is actually the density
function of a mixed discrete and continuous distribution \cite{Feller71}.
The probability of having a population of zero is 0.471,
which corresponds to the fraction of least-squares fits with an active
nonnegativity constraint on the gauche$^+$ gauche$^+$ population.
The continuous part of the probability density function has a population
mean of 0.062 and has a roughly exponential distribution.
The overall gauche$^+$ gauche$^+$ population mean is as expected near zero
because the two rotational isomer model, which generates
the Monte Carlo NMR observables, lacks gauche$^+$ gauche$^+$ rotational isomer.
The mean of the continuous part of the probability distribution
is greater than 5\% and suggests in perhaps a more direct fashion
than fitting the three rotational isomer model as discussed in the
last paragraph that gauche$^+$ gauche$^+$ populations in
the 5\% range can not be measured by fitting the NMR data with
currently available models.
The extent of the continuous part of the gauche$^+$ gauche$^+$
probability density function is determined by the observation errors
incorporated in the least-squares design matrix and observation vector.
Though this extent has nothing to do with the 2.1 kcal/mol
relative energy of the gauche$^+$ gauche$^+$ rotational
isomer determined by molecular mechanics or the accuracy of this energy,
it places gauche$^+$ gauche$^+$ population in a molecular mechanically
realistic range and approximately captures the uncertainty
in the molecular mechanics energy well depths.
In this sense the Monte Carlo procedure described here
blends together molecular mechanics and experimental NMR data.

\begin{figure}
\resizebox{8.3cm}{!}{\includegraphics{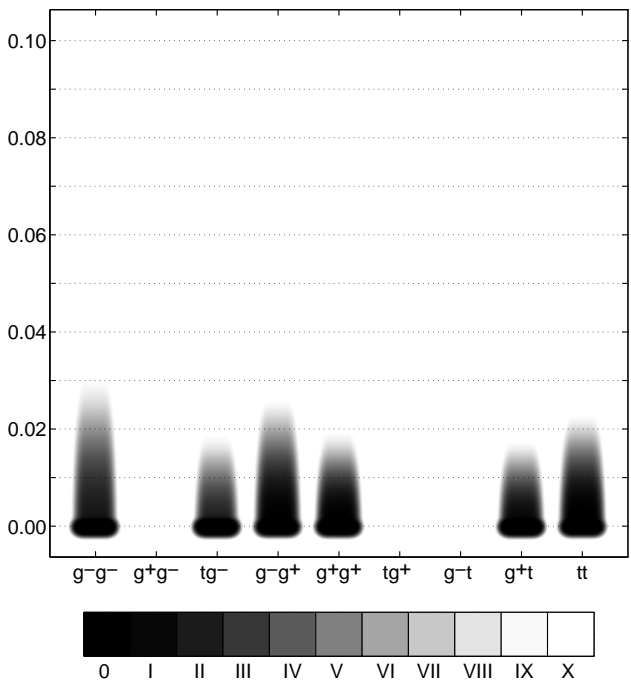}}
\caption{Gel graphic of rotational isomer population probabilities with 
Karplus equation coefficient and molecular mechanics geometry errors removed.
The least-squares fit has only NMR measurement errors.
Note the ten-fold expanded vertical scale, which places
the prominent trans gauche$^+$ and gauche$^-$ trans rotational
isomers off scale.
All other rotational isomers populations are fixed at zero.
Their probability density functions show the unmeasurable population ranges.
The forbidden gauche$^+$ gauche$^-$ rotational isomer is not included.}
\label{fig4:wish_likely_magnify}
\end{figure}

The observation errors of the least-squares fit are dominated
by the errors in the predicted NMR observables due to uncertainty
in the Karplus coefficients and molecular mechanics geometries.
What would happen if the errors in the predicted NMR observables
could be reduced below the level of the experimental measurement errors?
Given that the errors in the predicted NMR observables
are about an order of magnitude larger than the experimental
measurement errors, the population error estimates should also be
reduced by about an order of magnitude into the 0.5\% range.
A population of 0.5\% corresponds to a relative rotational
isomer energy of around 3 kcal/mol.
Excepting the prominent trans gauche$^+$ and gauche$^-$ trans
and the forbidden gauche$^+$ gauche$^-$ rotational isomers,
the remaining rotational isomers have energy map minima
ranging 2.1 kcal/mol for gauche$^+$ gauche$^+$
to 3.9 kcal/mol for gauche$^-$ gauche$^-$.
A 0.5\% population accuracy potentially places all rotational isomer
populations except that of the forbidden gauche$^+$ gauche$^-$
rotational isomer within reach of experimental measurement.
The measurability of the populations of all these rotational isomers
can be assessed by the same Monte Carlo procedure applied
in the previous paragraph to assess gauche$^+$ gauche$^+$ measurability
(Table \ref{tab3:ModelsSummary} row 38).
Again we fit the simple two rotational isomer model
with all observation errors included.
The Monte Carlo NMR observables are generated with only
the experimental measurement errors and an eight rotational
isomer model that excludes only the forbidden gauche$^+$ gauche$^-$
rotational isomer is then fit to the Monte Carlo observables.
In the initial model all rotational isomers except the
prominent trans gauche$^+$ and gauche$^-$ trans rotational isomers
are fixed at zero population.  As a result the Monte Carlo
probability densities (Fig.\ \ref{fig4:wish_likely_magnify})
of these rotational isomers are the density functions
of mixed discrete and continuous distributions
with zero population probabilities ranging from 0.52 to 0.89
and population means ranging from 0.0005 to 0.0023.
The Monte Carlo density functions give population estimates
of $0.3712 \pm 0.0038$ and $0.6203 \pm 0.0051$
for the trans gauche$^+$ and gauche$^-$ trans rotational isomers.
The continuous parts of the probability density functions
have population means ranging from 0.0035 to 0.0067
and have roughly exponential distributions.
This seems to confirm that if experimental measurement was the only
source of error, then rotational isomer populations as small
as about 0.5\% could be measured.

\section{CONCLUSIONS}

This study the cobalt glycyl-leucine dipeptide side chain rotational isomer
populations gives a realistic picture of their measurability
and in particular suggests that the population of the gauche$^+$ gauche$^+$
rotational isomer is less than 5 or 10\%, which is below the limit
of measurability at the present accuracy of Karplus equation calibration.
Better calibrations of the Karplus equations with model systems
such as cobalt dipeptides
promise to push the limit of measurability of side chain
populations down into the 1\% range.
To calibrate the Karplus equations to an accuracy substantially better
than $\pm$1 Hz probably requires a separate Karplus equation
for each vicinal coupling constant, for example, the \jj{H$\alpha$}{H$\beta$}
couplings of the leucine side chain would require calibrating
two Karplus equations, one for the coupling to the $\beta1$-proton
and a second for that to the $\beta2$-proton.  Each Karplus
equation has three coefficients, so the calibration of the
two \jj{H$\alpha$}{H$\beta$} Karplus equations could require
that as many as six coefficients be determined.
As many as four of these six coefficients could be determined from
the temperature dependence
of the two vicinal coupling constants \cite{Whitesides67}.
However, it is probably best to attempt to determine only three
of these coefficients by adopting the following compromise.
Because the torsion angle $\phi$ between the $\alpha$ and either $\beta$-proton
is always synclinal or antiperiplanar the Karplus equations
only need to be really accurate within these angular ranges.
Reasonable accuracy in these ranges can probably be achieved
with only two adjustable parameters per Karplus equation,
say by adjusting the $A$ and $C$ coefficients and fixing the coefficient
$B$ at the best literature value. Note that these coefficients
are defined in the background section and that $B=(\jjnm{ap}-\jjnm{pp})/2$,
where $\jjnm{pp}$ and $\jjnm{ap}$ are the coupling constants
at $\phi=0$ and 180 degrees.
A further reduction by one in the number of coefficients could
be achieved by assuming that $\jjnm{pp}$ is the same for both
$\beta$-protons.  With these compromises there are only three
coefficients to be determined from four measured quantities,
that is, two coupling constants and their temperature variations.
By measuring a half-dozen or so coupling constants about
the C$^{\alpha}$--C$^{\beta}$ and C$^{\beta}$--C$^{\gamma}$ bonds
of the leucine side chain there would be more than enough
measured values to calibrate all the Karplus equations
and determine the unknown rotational isomer populations.
Fitting the temperature dependence of the coupling constants
implies a search for not only the rotational isomer populations,
but also for a breakdown of the free energy differences
into enthalpic and entropic contributions.
The entropic contribution to the
the population differences can probably be estimated more
accurately from molecular mechanics \cite{Haydock93} than
by fitting the temperature dependence of the vicinal coupling constants.
Before attempting to calibrate Karplus equations with
the cobalt glycyl-leucine dipeptide
the critical issue of assigning the $\beta$ and $\delta$-protons
must be addressed by preparing samples with stereoselectively
deuterated leucine side chains \cite{Ostler93}.
The L-amino acid leucine stereoselectively deuterated
at the C$^{\delta}$ position is available from
Cambridge Isotope Laboratories, Inc., 50 Frontage Road, Andover MA 01810.
It is also important to measure the cobalt glycyl-leucine dipeptide
side chain rotational isomer populations with experiemnts that
are completely independent of any Karplus equation calibration.
Triple quantum filtered nuclear Overhauser effect spectroscopy (3QF-NOESY)
or tilted rotating frame Overhauser effect spectroscopy (3QF T-ROESY)
experiments give torsion restraints from cross correlated relaxation
\cite{Bruschweiler89}
and could supply this independent corroboration of the population estimates.

In addition to calibrating Karplus equations with the cobalt glycyl-leucine
dipeptide and other cobalt dipeptides we suggest parallel studies of
$N$-acetyl-{\footnotesize L}-leucine $N^{\prime}$-methylamide
(leucine dipeptide), which is the leucine analogue
of the alanine dipeptide \cite{Gresh98,Brooks93}.
The cobalt dipeptides achieve a single
backbone conformation by forming two approximately planar chelate rings.
This backbone conformation is uncommon in proteins and destablizes
the normal side chain conformational preferences of leucine.
The leucine dipeptide should eliminate this disadvantage
of the cobalt glycyl-leucine dipeptide without sacrificing the significant
advantage of a single backbone conformation.
Indeed the alanine dipeptide C$_{7\text{eq}}$ backbone
conformation predominates in weakly polar solvents \cite{Madison80}.
This strongly suggests that the leucine dipeptide backbone would adopt
the C$_{7\text{eq}}$ conformation in solvents such as
acetonitrile or chloroform, which would in turn strongly favor
the gauche$^-$ trans conformation of the leucine side chain.
Parallel studies of cobalt dipeptides and alanine dipeptide analogues
would also reveal the extent to which cobalt chelation influences
the vicinal coupling constants about the C$^{\alpha}$--C$^{\beta}$ bond.

A molecular graphic with multiple superimposed structures
is perhaps the most common format for reporting
the presence of multiple conformations in an NMR
or crystallographic protein structure.
It is very tempting to suppose that the structures displayed
in these molecular graphics are correct in every detail
and that structures or side chain conformations not present
in these molecular graphics are extremely unlikely.
Gel graphics such as those presented here for the leucine side chain
of the cobalt dipeptide could supplement common molecular graphics
and give a more realistic picture of protein conformational distributions.

\section{METHODS}

We adopt the convention that a side chain torsion
angle is gauche$^-$ in the range $-$120 to 0 degrees,
gauche$^+$ in the range 0 to 120 degrees,
and trans in the range $-$180 to $-$120 or 120 to 180 degrees.
Torsion angle definitions and atom names are specified by
the IUPAC-IUB conventions and nomenclature \cite{IUPAC70}.
The $\beta$ and $\delta$-protons are also identified
according to the Cahn--Ingold--Prelog nomenclature scheme \cite{Voet95}
for substituents on the C$^{\beta}$ and C$^{\gamma}$ prochiral centers.
The torsion angle between two atoms across a single bond is synclinal
in the range $-$90 to $-$30 or 30 to 90 degrees and antiperiplanar
in the range 150 to 180 or $-$180 to $-$150 degrees \cite{Klyne60}.
These last terms are very convenient for describing the geometry
of vicinally coupled spin systems.

\subsection{NMR experiments}

Barium[glycyl-L-leucinatonitrocobalt(III)] was prepared as previously
described \cite{Juranic93}.
NMR spectra were recorded at 500 MHz on a Bruker AMX-500 spectrometer.
Pure absorption 2D NOESY spectra were obtained by time-proportional
phase incrementation \cite{Marion83}
at mixing times of 50, 100, 200, 400, and 800 ms
and the cross relaxation rates were determined by one parameter
linear least-squares fits of the initial build-up
rates of the peak volumes \cite{Fejzo89}.
The standard deviations of the cross relaxation rates
were estimated from the one element least-squares moment matrices.
Cross relaxation rates within one standard deviation
of zero were set equal to zero.
Both homonuclear and heteronuclear vicinal coupling constants were
derived from 1D NMR spectra.
The homonuclear couplings were analyzed \cite{Castellano64,Ferguson64}
with the LAOCN-5 program (QCPE \#458).


\subsection{Simple models for preliminary $\beta$-proton assignment}

The goodnesss-of-fits of the two alternative assignments
were compared for two simple models for the experimental
coupling constants across the C$^{\alpha}$--C$^{\beta}$ bond.
The two rotational isomer model included only the gauche$^-$
and trans $\chi^1$ rotational isomers and the three rotational
isomer model included all three $\chi^1$ rotational isomers.
For both models the experimental data was linear least-squares fit
with the population sum constrained to one so that
the first model had one population parameter and the second two parameters.
The torsion angles between the coupled spins were assumed to have ideal
synclinal or antiperiplanar values of magnitude 60 or 180 degrees.
The Karplus coefficients for the \jj{H$\alpha$}{H$\beta$} and
\jj{C$^{\prime}$}{H$\beta$} coupling constants
(Table \ref{tab4:KarplusCoefficients})
were specifically calibrated \cite{Cung82,Fischman80} for these
couplings in peptides without any correction for cobalt chelation effects.
The predicted coupling constants were assumed to have errors of 1.5 Hz
to accommodate uncertainty in both the Karplus coefficients and geometry.

\subsection{Molecular mechanics}

Energy maps on $\chi^1 \times \chi^2$ torsion space and
cobalt glycyl-leucine structure coordinates were calculated with
the CHARMM molecular mechanics program \cite{Brooks83}.
The CHARMM structure file internal data structure was generated
from custom made topology and parameter CHARMM input files.
The custom topology file contained nonstandard glycine and leucine
residues to generate the peptide portion of the cobalt dipeptide complex
and a patch to add the cobalt and three nitro groups.
The molecular mechanics atomic charges were assigned
by a simple scheme.
First the side chain and backbone atomic charges
were set to those of a glycyl-leucine zwitterion.
The net dipeptide charge was then $-$0.6 because
one N-terminal amine proton and the backbone amide proton
were removed to make cobalt bonds.
On the assumption that the nitro group charge magnitudes should
be somewhat less than unity and the cobalt charge should
be slightly positive we then assigned a charge of $-$0.6 to each
nitro group and 0.4 to cobalt to give the correct total charge
of $-$2.0 to the complete cobalt dipeptide anion.
Electrostatics interactions were computed at a dielectric
constant of 1.0 and without any distance cutoff, except where otherwise noted.
Specialized force field parameters for the cobalt dipeptide
ring system were introduced by giving nonstandard atom
type codes to all the peptide backbone heavy atoms
except for the leucine $\alpha$-carbon atom.
Bond lengths and bond angles were taken
from a glycyl-glycine cobalt crystallographic structure \cite{Prelesnik84}.
Bond length and bond angle force constants were taken
from the force field of a polyamine cobalt complex \cite{Buckingham74}.
The force constants of torsion angles with cobalt
in one of the four angle defining positions were guessed by
adopting the torsion angle force constants of standard peptide
backbone atoms with roughly matching orbital hybridization and bonding.
A glycyl-glycine cobalt crystallographic structure \cite{Herak82}  
gave the initial Cartesian coordinates of the cobalt dipeptide
ring system.  Other ligand heavy atoms of this
same structure gave initial coordinates for the three nitro nitrogens.
The nitro oxygens were then built so that
the plane of each nitro group was in a staggered orientation
with respect to the other cobalt ligand bonds as viewed
down each nitro cobalt bond.
An internal coordinate representation of the leucine side
chain was setup during generation of the CHARMM structure file.
The torsion angle internal coordinates were set to ideal values
defined in the topology file and the bond lengths and angles
were filled in from the parameter file.
The molecular mechanics energy map over $\chi^1$ and $\chi^2$ torsion space
was computed by editing these two torsion angle
internal coordinates, building the
side chain Cartesian coordinates from internal coordinates,
restraining the $\chi^1$ and $\chi^2$ torsion angles
with an energy constant of 400 kcal/mole$\cdot$rad$^2$,
and energy minimizing by the steepest decent method for 20 steps followed
by the adopted basis Newton-Raphson \cite{Brooks83} method for 200 steps.
This sequence of edit, build, restrain, and minimize was
repeated for all 5184 points on a 5 degree grid
in $\chi^1$ and $\chi^2$ torsion space.
The map was output from CHARMM as a list with each line
containing the $\chi^1$ and $\chi^2$ coordinates and energy at one grid point.
The Cartesian coordinates of the energy minimized dipeptide
were temporarily output as a trajectory file
with one trajectory file coordinate set for each grid point.
Interatomic distances and torsion angles required for modeling
cross relaxation rates and vicinal coupling constants were
extracted from this temporary trajectory file with the CHARMM
correlation and time series analysis command.
To account for rotational averaging of the cross relaxation rate
to a methyl group, the three interatomic distances
from the methyl protons to the other cross relaxing atom
were inverse sixth root mean inverse sixth power
averaged with CHARMM time series manipulation commands.
Each interatomic distance, averaged interatomic distance, and vicinal
spin torsion angle was output from CHARMM as a separate file with
one distance or angle on each line and one line for each grid point.

Compared to the CHARMM 22 developmental topology and parameter files
for proteins with all hydrogens
our custom topology file has similar backbone charges
and side chain charges of about one half the magnitude.
Our peptide parameters are generally similar to those in the
CHARMM 22 developmental parameter file,
except that we only define a single tetrahedral
carbon atom type that does not depend on the number
of bonded hydrogen atoms and our bond angle potential
does not have Urey-Bradley \cite{Brooks88} interactions.

A separate FORTRAN program \cite{Haydock93} partitioned the molecular
mechanics $\chi^1 \times \chi^2$ energy map into energy well regions.
The program employed a cellular automata that adjusted the regions
so that the boundaries passed through the energy map saddle points
and followed along the ridges leading up to the tops of high energy peaks.
The program named each well, assigned an index to each well,
arranged the indices in a conventional order,
and output a new energy map such that each output line
specified the energy and well index of one grid point.
This cellular automata program is available from the
first author of this paper.

\subsection{NMR observables and Monte Carlo simulations}

We calculated NMR observables
(both NOESY cross relaxation rates and vicinal coupling constants),
fit rotational isomer probabilities,
ran Monte Carlo simulation of probability distributions, and generated
graphics with the MATLAB software package.
To accomplish these tasks we carefully designed and wrote a library
of 36 function files containing about 1600 lines of MATLAB script.
These functions passed all variables explicitly through input and output
argument lists and made no references to global variables except
in one minor instance of a function passed through an input argument list.
Important information, such as spin assignments, isomer names,
NMR measurement names, Karplus coefficient selections, and
$\chi^1 \times \chi^2$ torsion space grid point coordinates,
was passed explicitly from low level definition routines back
to high level I/O routines to minimize the possibility of
mixing up array index definitions.

The NMR observables were calculated by a function file that
input a list of NOESY cross relaxing protons and vicinally coupled spins,
opened appropriate CHARMM distance or angle data files,
which are described in the mechanics methods subsection,
and output a matrix of cross relaxation rates and vicinal coupling constants,
where the matrix had one column for each NMR observable and one row
for each distance or angle listed in the CHARMM files.
To calculate a cross relaxation rate
the molecular mechanics interproton distance read from the CHARMM data file
was raised to the inverse sixth power and multiplied
by the average of glycine geminal $\alpha$-proton
and leucine geminal $\beta$-proton scale factors,
where each scale factor was equal to the experimental geminal proton
cross relaxation rate times the sixth power of average molecular mechanics
distance between the geminal protons.
The glycine and leucine geminal proton relaxation rates were
$0.39 \pm 0.01$ s$^{-1}$ and $0.42 \pm 0.01$ s$^{-1}$ respectively.
The geminal proton scale factor varied from 13.5
to 13.7 \AA$^6$s$^{-1}$ depending on whether
the geminal proton distances were averaged over
structures at all energy map grid points or just over
the structures of the nine energy minimized rotational isomers.
To calculate a vicinal coupling constant the function file
selected the Karplus equation coefficients
based on the names in the input list of vicinally coupled spins and
inserted the molecular mechanics vicinal proton torsion angle read
from the CHARMM data file into a Karplus
equation with the selected coefficients.
The Karplus coefficients for the \jj{H$\alpha$}{H$\beta$}
and \jj{C$^{\prime}$}{H$\beta$} coupling constants
were specifically calibrated \cite{Cung82,Fischman80} for these
couplings in peptides without any correction for cobalt chelation effects.
The coefficients for the \jj{H$\beta$}{H$\gamma$}
coupling constant were those suggested \cite{Pachler72,BothnerBy65}
for the {\em sec}-butyl fragment without correction for the extra
carbon substitution on the $\gamma$-carbon.
The coefficients for the H$^{\alpha}$--C$^{\alpha}$--C$^{\beta}$--C$^{\gamma}$
and C$^{\alpha}$--C$^{\beta}$--C$^{\gamma}$--H$^{\gamma}$ heternuclear
coupling constants were taken from a fit to theoretical coupling constants
calculated for propane \cite{Breitmaier90,Wasylishen73,Wasylishen72}.
These Karplus coefficients are summarized in
Table \ref{tab4:KarplusCoefficients}.

The molecular mechanics energy, interproton distances, vicinal proton
torsion angles, and NMR observables were all computed on a 5 degree
$\chi^1 \times \chi^2$ torsion space grid.
The average $\chi^1$ and $\chi^2$ torsion angles and average
NMR observables of each rotational isomer were computed
by Boltzmann weighted summation over each energy well region
in $\chi^1 \times \chi^2$ torsion space.
To average the $\chi^1$ and $\chi^2$ torsion angles
over an energy well region these angles were
referenced to the minimum energy grid point
so that they varied continuously in the range
$-$180 to 180 degrees in this energy well region.
We assessed the effect of thermal motions by comparing the
average NMR observables of each rotational isomer with
the observables at the average $\chi^1 \times \chi^2$ torsion angles,
at the $\chi^1 \times \chi^2$ energy map minimum position,
and at the ideal geometry $\chi^1$ and $\chi^2$ torsion angles.
The NMR observables at the average torsion angles were computed
by interpolating the observables between $\chi^1 \times \chi^2$
torsion space grid points with a bicubic spline.
The energy map minima positions were approximated by the minima
of the interpolating function.
To assess the accuracy of this approximation we repeated the
molecular mechanics energy minimization at the minimum energy
grid point in each of the nine energy wells, except that during
the last 100 steps of adopted basis Newton-Raphson minimization
the $\chi^1$ and $\chi^2$ torsion angle restraints were released.
The $\chi^1$ and $\chi^2$ torsion angles of these minimized unrestrained
structures differed from the interpolated energy map minimum positions
by less than about one half degree for all
the energy wells except for the highly anharmonic
trans gauche$^-$ and trans trans energy wells, where
the torsion angles differed from the interpolated positions
by about one and one half degrees.
Given these small differences in $\chi^1$ and $\chi^2$ torsion angles
we assumed that the differences between the distances, angles,
and NMR observables interpolated to the energy map minima positions
and those calculated from the minimized unrestrained
structures would also be small.
We arbitrarily decided to calculate the NMR observables
at the $\chi^1 \times \chi^2$ energy map minimum positions
from the minimized unrestrained structures rather than calculate
them by interpolating the observables between $\chi^1 \times \chi^2$
torsion space grid points.

To find the rotational isomer probabilities
we minimized the difference between the experimentally measured
and predicted NOESY cross relaxation rates and vicinal coupling
constants subject to the constraints that the probabilities
were nonnegative and that their sum was one.
The design matrix was formed by calculating the NMR observables
for each rotational isomer as described in the last two paragraphs,
arranging these observables in a matrix with one row for each observable
and one column for each rotational isomer, and dividing each matrix
element in each row by the observation error for that row.
The observation vector was formed by dividing element-wise the column vector
of experimental measurements by the column vector of observation errors.
The observation errors were the RMS average of the experimental measurement
errors (Table \ref{tab2:NMRdata})
and errors in the predicted NMR observables due
to uncertainty in the Karplus coefficients and molecular mechanics geometries.
The predicted NOESY cross relaxation rates were assumed to have
uncorrelated errors of 0.01 s$^{-1}$ and the predicted
vicinal coupling constants were assumed to have uncorrelated errors of 1.0 Hz.
The linear least-squares with linear constraints problem
was converted to the equivalent \cite{Gill81} quadratic
programming problem and solved by an active set strategy \cite{Gill84}.

The accuracies of the fit rotational isomer probabilities
were estimated from Monte Carlo probability density functions
and from the diagonal elements of a moment matrix.
We setup an unconstrained linear least-squares subproblem with the number
of rotational isomer probability parameters equal
to one less than the number of inactive nonnegative probability constraints.
The desired moment matrix was obtained by transformation \cite{Cramer46}
of the subproblem moment matrix to generate matrix elements
for the probability parameter that was removed to enforce
the probability sum constraint.
The probability density functions of the fit rotational isomer probabilities
(parameters) were computed by the standard Monte Carlo recipe \cite{Press89}:
the experimental NMR observables were fit
to yield fit parameters and fit NMR observables,
the fit parameters and fit NMR observables were assumed
to be the true parameters and the error free experimental NMR observables,
random errors were added to the fit NMR observables to give
simulated NMR observables,
these simulated NMR observables were fit
to give simulated fit parameters, the previous two steps were
repeated may times and the resulting large set of simulated fit parameters
was histogramed to form the Monte Carlo probability density functions
of the fit parameters (rotational isomer probabilities).
To keep the simulated NMR observables nonnegative the random errors
were drawn from appropriately truncated Gaussian distributions.
These distributions were generated by a simple acceptance-rejection method,
that is, if one sample drawn from a standard Gaussian distribution would
have given a negative value to a particular NMR observable then that sample
was discarded and a new sample was drawn and tested in the same way.
When the standard deviations of the Monte Carlo probability density functions
were significantly smaller than the fit rotational isomer probabilities,
that is, when none of the fit rotational isomer
probabilities were near zero, the Monte Carlo standard deviations were
almost exactly equal to the moment matrix standard deviations.

\subsection{Gel graphics}

Monte Carlo probability density functions were
displayed as gel graphics, which were designed to visually indicate both
the discrete probability fraction at zero population and shape of the
continuous probability density over the range of population from zero and one.
This was accomplished by a simulated photographic process where the degree
of film overexposure indicates the probability fraction at zero population
and continuous gray tones represent the continuous part of the probability
density.  The continuous part of the probability density
was prefiltered to reduce the noise from the Monte Carlo sampling.
The prefilter consisted of two cycles of alternating extrapolation and
Gaussian smoothing.  The first cycle estimated the probability density
at zero population by evenly extrapolating the probability density
about zero population and then smoothing.  In the second cycle the
original probability density was oddly extrapolated about
the just determined probability density at zero population and then smoothed.
The standard deviation of the smoothing in the first cycle was
half that in the second cycle.  We speak of this second cycle standard
deviation as the prefilter standard deviation.
This somewhat cumbersome prefilter procedure smoothed
the continuous part of the probability density without introducing distortion
near zero population, where the probability density typically
has a positive value and a nonzero slope.  Note that this positive value
is distinct from the discrete probability fraction at zero population,
which is not prefiltered.  The prefiltering eliminated
noise from the Monte Carlo sampling, which would otherwise
show up as distracting transverse stripes across the gel lanes.
We examined the convergence of the probability density functions
by repeating Monte Carlo simulations with increasing numbers of steps
and comparing conventional probability density plots and gel graphics.
The prefilter standard deviation was adjusted to remove Monte Carlo
sampling noise without visibly obscuring the shape of the probability density.
For simulations of length $10^3$, $10^4$, and $10^5$ steps the appropriate
standard deviation expressed as the full width at half max (FWHM)
was 32, 16, and 8 histogram bins, where 1024 bins covered
the population range from zero to one.
At $10^3$ steps substantial fluctuations in the shape of the conventional
plots were observed, but the gel graphics had at least qualitatively converged
to their final appearance.  At $10^5$ steps only tiny differences
were observed in the conventional plots and the differences in the
gel graphics were imperceptible.  The gel graphics presented here
were created from Monte Carlo simulations of $10^5$ steps, even though
simulations as short as $10^3$ steps would be adequate for many purposes.

After prefiltering an initial image was formed with
the probability density functions of the rotational isomers
displayed in lanes 1024 rows long by 64 columns wide.
In this initial image the values across each row
were constant and equal to the probability of finding
the population in bins of width $1/1024$ covering the range zero to one,
except that the values across bottom row of each lane contained the discrete
probability fraction at zero population.
The probability of finding the population in a given bin is proportional
to the bin width and inversely proportional to the total number of bins,
but the probability fraction at zero population is independent of the bin width.
Because of the relatively high resolution of the image
the probability fraction at zero population is about an order
of magnitude larger than the probability of finding the population
in another bin.  For this reason the probability fraction at zero population
can be effectively displayed as a film overexposure.

To simulate film overexposure at zero population and smooth the lane edges
along the continuous part of the probability distribution a Gaussian blur
filter with a FWHM of 16 pixels was applied to the initial image.
With this amount of blurring the typical probability density
at zero population was still considerably greater than
that along the continuous part of the probability distribution.
The pixel values of the blurred image were treated like scene
luminances \cite{Zalewski95} and converted into photographic
print densities.  First, the maximum printable luminance $L_m$ was
set equal to the maximum probability in the continuous part
of the probability distributions, that is, excluding the probability
fractions at zero population.  The logarithm of the luminance $L$
was converted to density with a characteristic curve \cite{Hunt95}
given by $(3x^2-2x^3)10\log2$, where
$x=1+(\log{L}-\log{L_m})/(10\log2)$ and $0 \le x \le 1$.
Note that the maximum point-gamma of our characteristic curve is 1.5.
Then the printable densities were linearly mapped into gray scale values.
A stepwedge bar of the 11 zones in the Zone System \cite{Adams95}
was added to the gel graphic as an aid to calibrating
the probability densities.


\subsection{Supporting information available}

Molecular mechanics, data analysis, and gel graphics input files
and additional data tables\footnote{The supporting information includes
CHARMM input files to generate energy minimized
rotational isomers without and with torsion restraints
and to extract the vicinal coupling torsion angles
and NOESY interatomic distances,
crystallographic coordinates of cobalt dipeptide chelate rings
with added nitro groups,
cobalt glycyl-leucine dipeptide topology and parameter files,
MATLAB version 4 M-files to calculate NMR observables, fit rotational isomer
probabilities, simulate Monte Carlo probability distributions,
and to generate gel graphics,
tables of torsion angles and interatomic distances of the energy minimized
rotational isomers without torsion restraints
and of vicinal coupling constants and NOESY cross
relaxation rates calculated at these angles and distances.},
which are sufficient to reproduce all the results reported here,
are included in this electronic preprint's source file.

\section*{ACKNOWLEDGMENTS}

We thank Martin Karplus for access to the c22g2 release of the
CHARMM 22 program system and thank Roland L. Dunbrack, Jr.\
for providing the torsion angle list for constructing the
region-of-interest rotamer library\footnote{The torsion
angle list can be downloaded
from the backbone-dependent rotamer library Web page
through a hyperlink to the file of dihedral angles for all chains
at http://www.fccc.edu/research/labs/dunbrack/sidechain.html .}.
We are grateful to the National Institutes of Health (GM 34847)
for financial support of this study.

\bibliography{dip}

\end{document}